\def\DThirtyNineNamedDraft{1}
\documentclass[10pt]{article}
\ifdefined\DThirtyNineNamedDraft
\usepackage[preprint]{tmlr}
\else
\usepackage{tmlr}
\fi

\usepackage[utf8]{inputenc}
\usepackage{amsmath,amssymb,amsfonts,bm}
\usepackage{booktabs,tabularx,array,multirow}
\usepackage{graphicx,xcolor}
\usepackage{caption,subcaption}
\usepackage{float}
\usepackage{needspace}
\usepackage[section]{placeins}
\usepackage{enumitem}
\usepackage{hyperref}
\usepackage{url}
\usepackage[capitalise,noabbrev]{cleveref}

\hypersetup{colorlinks=true,linkcolor=blue!55!black,citecolor=blue!55!black,urlcolor=blue!55!black}
\setlist{itemsep=1pt,topsep=3pt}


\newcommand{\PiF}{\Pi_{\mathcal F}}
\renewcommand{\Re}{\mathrm{Re}}
\newcommand{\Evol}{E_{\mathrm{vol}}}
\newcommand{\Enear}{E_{\mathrm{near}}}
\newcommand{\Egrad}{E_{\nabla}}
\newcommand{\Epk}{E_{\mathrm{pk}}}
\newcommand{\Eobj}{E_{\mathrm{obj}}}
\newcommand{\Robj}{R_{\mathrm{obj}}}
\newcommand{\status}[1]{\textsc{#1}}

\newcommand{\HsixItMin}{3{,}460}
\newcommand{\HsixItMax}{7{,}245}
\newcommand{\HsixDiscGlobal}{1.7\times10^{-5}}
\newcommand{\HsixDiscNear}{3.8\times10^{-5}}
\newcommand{\HsixDrift}{3.6\times10^{-8}}
\newcommand{\HsixMetric}{2.0\times10^{-5}}
\newcommand{\HsixBpTenLo}{1.9\times10^{-7}}
\newcommand{\HsixBpTenHi}{7.0\times10^{-7}}
\newcommand{\HsixBpNineLo}{2.0\times10^{-3}}
\newcommand{\HsixBpNineHi}{3.3\times10^{-3}}

\newcommand{\HsixScope}{all 12 cases}
\newcommand{\HsixGuard}{, and every predeclared iterative-sensitivity guard passes}

\title{Where the Error Lives:\\ Separating Lid-Edge and Body-Induced Error\\ in Neural Pressure Prediction for 3D Cavity Flow}

\ifdefined\DThirtyNineNamedDraft
\author{%
\name Zirun Zhou$^{1,2,*}$, Zhi-Qian Zhang$^{2,\dagger}$, Yixuan Hong$^{3}$,\\
Guang Liu$^{4}$, Zhigang Liu$^{2}$, Lili Zhang$^{2}$\\[0.5em]
{\normalfont\small
$^{1}$School of Computing, National University of Singapore, Singapore\\[0.3em]
$^{2}$EXDYNA Pte. Ltd., 81 Ayer Rajah Crescent, \#02-49, Singapore 139967\\[0.3em]
$^{3}$Tandon School of Engineering, New York University, Brooklyn, NY, USA\\
\texttt{yh6392@nyu.edu}\\[0.3em]
$^{4}$Institute of Advanced Intelligence and Computing (IAIC),\\
Agency for Science, Technology, and Research (A*STAR), Singapore, 138632\\
\texttt{liu\_guang@a-star.edu.sg}}\\[0.5em]
{\normalfont\small $^{*}$Corresponding author: Zirun Zhou, \texttt{zirun.zhou@u.nus.edu}.\\
Alternate contact: \texttt{hhvrfn@outlook.com}.\\
$^{\dagger}$Co-corresponding author: Zhi-Qian Zhang, \texttt{z@exdyna.ai}.\\
Alternate contact: \texttt{zhangzhiqian@meshfree.org}.}%
}
\hypersetup{pdfauthor={Zirun Zhou, Zhi-Qian Zhang, Yixuan Hong, Guang Liu, Zhigang Liu, Lili Zhang}}

\else
\author{\name Anonymous authors}
\fi

\begin{document}
\maketitle
\begingroup
\renewcommand{\thefootnote}{}
\footnotetext{\textbf{AI assistance.} OpenAI ChatGPT/Codex and Anthropic Claude tools assisted with research planning, code development, analysis, figure preparation, drafting and revision. Reported numerical results come from executed computations, not model-generated assertions. Responsibility for the work remains with the human authors.}
\endgroup

\begin{abstract}
Can a pressure surrogate score much better globally without better resolving the pressure around immersed bodies? We test this in a three-dimensional lid-driven cavity with an unregularized lid and centered pressure at native $128^3$ resolution. On 60 cases from 12 unseen geometries, with predictions sealed before the CFD truth existed, adding a body-free, Reynolds-conditioned reference correction to the same frozen $64^3$ surrogate lifted to $128^3$ lowers the mean full-volume relative $L_2$ error from 20.6\% to 7.5\%. In the post hoc regional analysis, the lid-edge junction occupies 1.5\% of the fluid but holds 69\% of the truth pressure energy and 95\% of the uncorrected surrogate's squared error. Interior error changes by just $+0.16\%$ relative, and object-local peak error and recall change little on average. Retrospective ablations locate the learned term's contribution in the interior and near the bodies; rebasing on the reference lowers global error further with little change in cohort-mean near-body metrics. Reference-relative body-response error remains about 25\% in the interior near-body band. Grid-refinement, lid-regularization and convergence checks delimit the numerical scope. In this benchmark, a large global gain chiefly reflects correction of a shared boundary field, not better resolution of the body response. We recommend reporting regional truth-energy and model-error shares alongside near-body and reference-relative body-response errors.
\end{abstract}

\section{Introduction}\label{sec:intro}

Learned surrogates model flow in complex three-dimensional domains \citep{gino,geo3d}; volumetric pressure prediction is one such task \citep{aortic3d}, and FlowBench provides volumetric pressure data \citep{flowbench}. Their accuracy is usually summarized by one relative $L_2$ error over the fluid volume, sometimes with an error near the body. Such a summary does not, by itself, establish how accurately a surrogate captures the pressure response near immersed bodies. Evaluation practice in scientific machine learning has been criticized for weak baselines and optimistic reporting \citep{mcgreivy2024}, and aggregate metrics are known to hide failures on important subsets \citep{hiddenstrat}. Here we isolate a specific source of this evaluation gap: for a transferred coarse pressure surrogate, a geometry-shared lid-edge field dominates the global error, while accuracy near the bodies follows a different pattern. Our main contribution is to locate and distinguish these errors. A reference correction serves as a controlled intervention that makes the distinction measurable, rather than as evidence that a lower global score necessarily means better body-conditioned predictions.

\begin{figure}[!htb]
\centering
\includegraphics{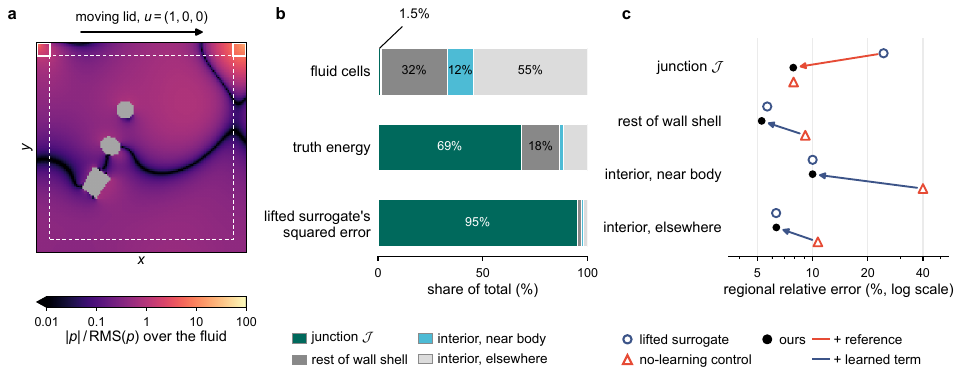}
\caption{\textbf{Where the error lives (cohort A, 60 native $128^3$ cases).} \textbf{a}, Centered CFD pressure magnitude relative to the case's fluid RMS on the $xy$ plane $z=43.5/128$ of cohort-A heterogeneous triplet-03 at $\Re=300$ (log scale; grey, solid). The lid moves in $+x$. White boxes mark the junction cells $\mathcal J$ where this plane crosses the two lid edges perpendicular to the motion; the dashed line bounds the 8-cell wall shell. Black curves are sign changes of the centered pressure, where $|p|$ falls below 1\% of the RMS. \textbf{b}, Shares of fluid cells, truth pressure energy and the lifted surrogate's squared error by region (case means; segments below 12\% are unlabelled except the 1.5\% junction cell share; values in \cref{tab:regional}). \textbf{c}, Regional relative error of three compositions. A marker's colour names the component it contains (blue, learned; red, reference); an arrow's colour names the component added on the way to ours (black dot). Where no arrow is drawn, the component changes that regional error by less than a marker width. All regional results are \status{post hoc}; the no-learning control is \status{retrospective}. Region definitions are in \cref{sec:regions}.}
\label{fig:thesis}
\end{figure}

The setting is the lid-driven cavity with immersed obstacles, used as a benchmark for geometry-conditioned surrogates \citep{flowbench,geo3d,cfdbench}. Where the moving lid meets a stationary wall perpendicular to its motion, the velocity boundary condition is discontinuous, and the continuum pressure is singular \citep{taylor1962,shankar2000,botella1998}. On a finite-volume grid this singularity appears as large pressures in the few cells along those lid edges. \Cref{fig:thesis} shows the consequence for native $128^3$ reference fields. The lid-edge \emph{junction}, 1.5\% of the fluid cells, carries 69\% of the centered pressure energy. A surrogate trained at $64^3$ and transferred to the native grid does not fully reproduce this structure, and 95\% of its squared error sits there.

The reference correction lets us distinguish a boundary-driven global gain from the accuracy supplied by learning near the bodies. The chronology matters: the work began with a method, and the regional analysis followed. During development we found that the coarse-to-fine error of a $64^3$ surrogate concentrates near the walls (\cref{app:lineage}). We therefore compose a \emph{frozen} learned $64^3$ surrogate with the part of a Reynolds-conditioned empty-cavity pressure that block averaging to $64^3$ discards. The reference is fitted offline to 17 native-grid simulations, and the composition needs no retraining and no CFD at query time (\cref{sec:method}). Only after the prospective test did we decompose its results by region. The paper's principal contribution is this evaluation finding; the controlled intervention, ablations and numerical checks provide distinct layers of evidence:
\begin{enumerate}
\item \textbf{Principal finding: regional evaluation changes the interpretation.} In this configuration, both truth energy and the transferred coarse surrogate's squared error concentrate at the lid edges (\cref{sec:where}). Their shares are measured separately: large truth-energy weight alone does not force a model's error to concentrate there. The regional decomposition of $\Evol^2$ explains why global ranking can obscure body-conditioned accuracy and motivates region-resolved reporting (\cref{eq:decomposition}).
\item \textbf{Controlled intervention: a prospective reference comparison.} Adding the body-free reference to the frozen surrogate substantially lowers global, gradient and peak-region errors on 60 cases from 12 unseen geometries (\cref{sec:prospective}). The regional analysis locates that gain rather than interpreting it as new local capability. The second cohort provides a \status{post hoc} reconstructed contrast. A retrospective rebasing baseline does better on every global metric, so the correction is not claimed to be optimal (\cref{sec:rebase}).
\item \textbf{Supporting ablations: complementary roles and a local exception.} Removing only the learned term identifies its substantial interior and near-body contribution (\cref{sec:learned}). One of 24 geometries across the two cohorts materially exceeds the local tolerances when the reference is added: its near-body band reaches furthest into the lid-edge junction (\cref{sec:boundary}). High-Reynolds-number object-local errors remain unresolved.
\item \textbf{Supporting numerical checks: scope and limitations.} Empty-cavity solutions at $64^3$, $128^3$ and $256^3$, a regularized-lid cohort and more tightly converged test-truth solutions separate grid, lid-profile and iterative-convergence effects (\cref{sec:physics}). These qualify the diagnostic finding rather than establish universal portability of the correction.
\end{enumerate}

\paragraph{Evidence status.} We label every quantitative result by its relation to the test data:
\begin{itemize}
\item \status{prospective}: predictions sealed with hashes, together with the evaluator, before the CFD truth existed;
\item \status{post hoc}: re-analyses of such frozen predictions after the truth was opened, for example by region;
\item \status{retrospective}: arms or retrainings created after the truth had been inspected.
\end{itemize}
Only two results are prospective. The first is the comparison of the lifted surrogate with the reference-corrected predictor on cohort A. The second is the absolute scores of the reference-corrected predictor on cohort B. The protocol follows the logic of preregistration \citep{nosek2018}.

\paragraph{Scope.} Inputs are geometry and a scalar Reynolds number. The output is kinematic pressure over the fluid volume, centered on the fluid mean, compared against a fixed OpenFOAM discretization at $128^3$. We do not predict surface pressure coefficients, forces or absolute pressure. Our reference fields are numerical solutions under one discretization, not continuum truth.

\section{Related work}\label{sec:related}

\paragraph{Geometry-conditioned pressure fields.} Volumetric pressure prediction from geometry has been studied for aortic and vascular flow \citep{aortic3d,vascular3d}, parameterized vehicle and exterior flows \citep{heger3d,geoformer,abupt}, turbomachinery \citep{blade3dno} and immersed swimmers \citep{swimmer}. DrivAerNet++ provides volumetric flow and surface-field data, with surface-pressure benchmarks \citep{drivaernet}. Operator-learning architectures address boundary--interior communication and irregular domains \citep{gino,bano,beno,dafno,domino,gaot,mno,globe}. None of these results is directly comparable to ours: domains, boundary conditions, pressure gauges and splits differ, and we compare only with baselines trained under our contract.

\paragraph{Benchmarks and evaluation.} FlowBench uses signed-distance conditioning for flow around objects in a lid-driven cavity and includes near-body evaluation \citep{flowbench}. Geometric-DeepONet uses its three-dimensional data and global/near-body metrics, but predicts velocity only in the cited version, despite pressure being present in the dataset \citep{geo3d}. Related cavity pressure tasks appear in CFDBench, ArGEnT, OTNO and a journal benchmark \citep{cfdbench,argent,otno,geometrymatters}. Broader benchmarks report boundary or frequency-resolved errors \citep{pdebench,thewell}, GeoABC distinguishes a region's error reduction from its contribution to the global reduction \citep{geoabc}, AirfRANS shows that global field errors can disagree with force-level accuracy \citep{airfrans}, and pressure-channel relative errors can substantially exceed velocity errors under a shared training objective in airfoil flow \citep{thuerey2020}. PINNacle regularizes the cavity lid so that its velocity vanishes at the corners \citep{pinnacle}. In a two-dimensional deep cavity with a regularized lid, \citet{roohi2026} show that the ranking of PINN training workflows varies across pressure-evaluation regions. In climate modeling, \citet{kwa2023} show that a state-independent correction improves global precipitation RMSE mainly over ocean while slightly worsening it over land. We test a different intervention: with the learned surrogate held fixed, we add a geometry-shared lid-edge pressure correction tied to one solver and grid pair and measure its effect separately from the body response. We do not re-analyse other datasets; whether they regularize the lid, and how strongly, determines how far our numbers transfer.

\paragraph{The lid-edge singularity.} Where a moving wall meets a stationary one at a right angle, the Stokes pressure behaves like $\nu U/r$ at distance $r$ from the edge (the scraper problem; \citealp{taylor1962,moffatt1964,shankar2000}). Inertia modifies this structure beyond $r\sim\nu/U$ \citep{hancock1981}. At $128^3$ that length is about 13 cells at $\Re=10$ and less than one cell at $\Re\geq300$, so the grid sees an inertia-modified corner at higher Reynolds numbers. Spectral benchmark solutions subtract the leading singular component analytically \citep{botella1998}, three-dimensional reference solutions treat the edges with care \citep{albensoeder2005,kuhlmann2019}, and some direct and large-eddy simulations regularize the lid velocity because the singularity degrades convergence near the lid \citep{leriche2000,bouffanais2007}. Singular functions have also been built into physics-informed networks for corner singularities of the Poisson problem \citep{sepinn}. Our reference correction plays a comparable role, but it is data-fitted and consistent with the solver's own discretization. It removes the grid-level representation of the singularity, not the continuum singularity.

\paragraph{Reference-informed learning and resolution transfer.} Reference Neural Operators learn a correction to a transported reference solution and note that the nearly constant density over most of the Airfoil-Euler domain can make whole-domain errors look small \citep{rno}. Multifidelity DeepONet learns a high-fidelity residual on top of a low-fidelity operator \citep{mfdeeponet}. Solver-in-the-loop training corrects a coarse numerical solver with a learned term \citep{um2020}, CFD-GCN refines an upsampled coarse-mesh CFD solution, including pressure, with a graph network \citep{cfdgcn}, and \citet{hanna2020} learn the local error of coarse-grid OpenFOAM solutions of the three-dimensional lid-driven cubic cavity. Coarse-to-fine correction in this very configuration is therefore not new; in all of these the correction is learned. Hybrid interpolation embeds CFD snapshots in a neural model \citep{shaawind}. We instead freeze an immersed-geometry surrogate and add a deterministic residual from a body-free library; \cref{tab:related} compares the information each approach uses. Zero-shot super-resolution of learned operators is unreliable \citep{falsesr,functionspace,bartolucci2023}, and directly executing our coarse surrogate at $128^3$ indeed degrades accuracy (\cref{app:lineage}). We therefore keep the learned field at its native $64^3$ scale.

\begin{table}[!htb]
\centering\small
\caption{\textbf{Information used by reference-informed approaches.} These are methodological distinctions, not numerical comparisons.}
\label{tab:related}
\begin{tabularx}{\linewidth}{@{}>{\raggedright\arraybackslash}p{38mm}>{\raggedright\arraybackslash}X>{\raggedright\arraybackslash}X@{}}
\toprule
Method & Solution information available & Learned vs.\ deterministic operations \\
\midrule
Reference Neural Operators \citep{rno} & reference-domain solution, reference and query geometry, a constructed deformation & transport the reference; learn the query-minus-transported correction \\
\addlinespace
Multifidelity DeepONet \citep{mfdeeponet} & low- and high-fidelity training data; low-fidelity predictions as inputs & learn a low-fidelity operator and a high-fidelity residual \\
\addlinespace
Hybrid wind interpolation \citep{shaawind} & CFD snapshots for a fixed object and sampled inlet angles & interpolation ansatz inside a trained neural model \\
\addlinespace
Ours & immersed-geometry training fields; an offline body-free library; query geometry and $\Re$ only at inference & freeze the surrogate; add the deterministic residual $(I-LR_2)B_{128}$; center \\
\bottomrule
\end{tabularx}
\end{table}

\section{Problem, data and evaluation}\label{sec:problem}

\FloatBarrier
\subsection{Flow configuration}
Reference fields are steady, incompressible, laminar solutions of OpenFOAM v2412 \texttt{simpleFoam} \citep{openfoam} in the unit cube. The plane $y=1$ moves with velocity $(1,0,0)$, and the other walls and the immersed bodies are no-slip. The kinematic viscosity is $\nu=1/\Re$ with $\Re\in\{10,30,100,300,900\}$. On a uniform $n^3$ grid ($n=64$ or $128$), cells whose centers have signed distance $\leq0$ to the body are removed, so the immersed wall is voxelized. All test cases and all 17 reference-library cases use byte-identical discretization and linear-solver settings (\cref{app:cfd}). Convergence requires residuals of $10^{-6}$ (pressure) and $10^{-7}$ (velocity) within 16{,}000 SIMPLE iterations. Pressure is kinematic and, since the lid speed and box size are one, it is measured in units of $U^2$. All fields are centered on the fluid mean:
\begin{equation}\label{eq:center}
\PiF q = m\Big(q - \tfrac{1}{|\mathcal F|}\textstyle\sum_{i\in\mathcal F} q_i\Big),\qquad m=\mathbf 1_{\mathcal F}.
\end{equation}

\FloatBarrier
\subsection{Cohorts}
The coarse surrogate is trained on 1{,}620 $64^3$ cases from 324 geometries of six families (sphere, cylinder, box, ellipsoid, torus, multi-body unions of two spheres) at the five Reynolds numbers. We evaluate at $128^3$ on two disjoint test cohorts of three shape or composition categories unseen in training: capsules, heterogeneous pairs (sphere + box) and heterogeneous triplets (sphere + cylinder + box). Each family has four geometries, one per placement stratum (interior; near $z{=}1$; near a random $x$ or $y$ face; near $z{=}0$), evaluated at all five Reynolds numbers. Each cohort therefore has 60 cases and 12 independent geometry groups.
\begin{itemize}
\item \textbf{Cohort A (primary).} Both the lifted surrogate and the reference-corrected predictor, the evaluator and the numerical criteria were sealed with hashes before the cohort's CFD truth was generated.
\item \textbf{Cohort B (second).} This cohort was generated later, with a different generator seed, to test a different model. The reference-corrected predictor was that test's sealed control arm, so its absolute scores on cohort B are prospective. The lifted arm was not sealed there. We reconstruct it deterministically from the sealed array by removing the correction (\cref{eq:intervention}), which makes the cohort-B contrast \status{post hoc}. Re-evaluating the sealed reference-corrected arrays reproduces its frozen cohort-B scores to $6\times10^{-17}$.
\end{itemize}
Coarse-grid cohorts used during development are described in \cref{app:lineage} and are not pooled with A or B.

\FloatBarrier
\subsection{Metrics and regions}\label{sec:regions}
For a region $\mathcal R$, prediction and truth are centered independently on the same fluid mask and compared by
\begin{equation}\label{eq:region}
E_{\mathcal R}(\hat p, p)=\|\PiF\hat p-\PiF p\|_{2,\mathcal R}\,\big/\,\|\PiF p\|_{2,\mathcal R}.
\end{equation}
\Cref{tab:metrics} lists the six case metrics predeclared for cohort A; lower is better except for the recall $\Robj$.

\begin{table}[!htb]
\centering\small
\caption{\textbf{Predeclared case metrics} (details and tie handling in \cref{app:metric}). All are dimensionless. The near-body band $\mathcal N=\{0<\mathrm{SDF}\le 0.2\}$ is 25.6 cells thick at $128^3$ and overlaps the wall shell for most bodies; only $\Eobj$ and $\Robj$ exclude cells at the outer walls.}
\label{tab:metrics}
\begin{tabularx}{\linewidth}{@{}lX@{}}
\toprule
Metric & Definition \\
\midrule
$\Evol$ & \cref{eq:region} on the full fluid volume $\mathcal F$ \\
$\Enear$ & \cref{eq:region} on the near-body band $\mathcal N$ \\
$\Egrad$ & Relative error of axiswise central differences where the center and both neighbours are fluid \\
$\Epk$ & \cref{eq:region} on the truth's top-5\% $|p|$ cells of $\mathcal F$ (global peaks) \\
$\Eobj$ & \cref{eq:region} on the truth's top-5\% $|p|$ cells of $\mathcal N$, excluding three cells at the outer walls (object peaks) \\
$\Robj$ & Fraction of those truth object-peak cells also in the prediction's same-size top-5\% set \\
\bottomrule
\end{tabularx}
\end{table}

\paragraph{Regions.} For the spatial analysis we partition the fluid by cell distance to the box faces:
\begin{itemize}
\item the \emph{wall shell} $\mathcal S$: cells within eight cells of any face;
\item the \emph{interior} $\mathcal I=\mathcal F\setminus\mathcal S$;
\item the \emph{lid-edge junction} $\mathcal J\subset\mathcal S$: cells within eight cells of both the lid and a stationary side wall. $\mathcal J$ has 30{,}720 cells (1.5\% of the fluid): two blocks along the edges perpendicular to the lid motion, two along the parallel edges, and four vertex blocks.
\end{itemize}
The 8-cell width comes from a development diagnostic that predates cohort A (\cref{app:lineage}). The split of the shell into junction and remaining wall classes was defined after cohort A had been evaluated, so all regional results are \status{post hoc}; \cref{tab:width} shows that the conclusions hold for junction widths from 1 to 16 cells. We also bin cells by Chebyshev cell distance to the nearest lid edge, i.e.\ $\max(d_{\mathrm{lid}}, d_{\mathrm{side}})$ minimized over the side walls, and, within $\mathcal I$, by distance to the body.

\paragraph{Reference-relative body response.} When an empty-cavity reference $B_{128}(\Re)$ is available, define
\begin{equation}\label{eq:body-response}
d=\PiF[p-B_{128}],\qquad \hat d=\PiF[\hat p-B_{128}],\qquad
E_{\mathrm{body},\mathcal R}=\frac{\|\hat d-d\|_{2,\mathcal R}}{\|d\|_{2,\mathcal R}}.
\end{equation}
Both disturbances use the same reference, query fluid mask, region and full-fluid gauge. The reference is body-free before masking and centering on the query fluid. Since $\hat d-d=\PiF\hat p-\PiF p$, only the denominator differs from \cref{eq:region}; this diagnostic is not independent evidence of accuracy or a decomposition into independent Navier--Stokes solutions. We average defined per-case ratios, not pooled squared norms. Zero disturbance energy makes the ratio undefined. Small positive energy can amplify relative errors, so we also report $\|d\|^2_{2,\mathcal R}/\|\PiF p\|^2_{2,\mathcal R}$ rather than clipping the score. The historical implementation used a $10^{-300}$ safeguard on squared denominator energy; it was inactive in all 485 recorded cohort-A regions with nonzero truth energy (minimum disturbance energy $0.0152\,U^4$).

\paragraph{Energy decomposition.} ``Pressure energy'' here abbreviates the squared $L_2$ norm of centered pressure, not physical fluid energy. Let $s_{\mathcal R}=\|\PiF p\|^2_{2,\mathcal R}/\|\PiF p\|^2_{2,\mathcal F}$ be a region's share of that norm. For any exhaustive partition, with the common full-fluid centering,
\begin{equation}\label{eq:decomposition}
\Evol^2=\textstyle\sum_{\mathcal R}s_{\mathcal R}\,E_{\mathcal R}^2 .
\end{equation}
The weights $s_{\mathcal R}$ are independent of the model, but its error-energy shares $w_{\mathcal R}=s_{\mathcal R}E_{\mathcal R}^2/\Evol^2$ are not (for $\Evol>0$). Thus large junction truth energy does not imply that every predictor's error is concentrated there. With $s_{\mathcal J}\approx0.69$ and $s_{\mathcal I}\approx0.13$ (\cref{tab:regional}), an interior error of 10\% contributes at most $\sqrt{0.13}\times10\%\approx3.7\%$ to $\Evol$.

\paragraph{Statistics.} The geometry is the independent unit ($n=12$ per cohort), and the five Reynolds numbers of a geometry are repeated measures. We report:
\begin{itemize}
\item cohort-mean error changes, i.e.\ the ratio of arithmetic means; recall uses the mean paired geometry-level difference;
\item 95\% $t$ intervals for the geometric-mean error ratio over geometries, a different estimand, so an interval need not contain the change of means; recall instead uses paired geometry-level differences and their $t$ interval;
\item per-geometry ranges and leave-one-geometry-out ranges;
\item two-sided sign-flip tests enumerating all $2^{12}$ patterns of geometry means. Enumeration is exact; inferential validity additionally requires sign symmetry under the null, not merely independent geometries.
\end{itemize}
For cohort A, the following \emph{decision rule} was predeclared (\cref{app:criteria}): at least 10\% mean reduction of $\Evol$, $\Egrad$ and $\Epk$; at most 2\% mean increase of $\Enear$ and $\Eobj$; at most 0.01 loss in $\Robj$. Reading those tolerances as non-inferiority margins is our post hoc choice. The resulting log-ratio intervals concern geometric-mean ratios, not the predeclared arithmetic-mean decision rule; they do not certify non-inferiority for that rule. Everything else is descriptive.

\section{Method: a frozen surrogate plus a body-free native-grid reference}\label{sec:method}

\begin{figure}[!htb]
\centering
\includegraphics{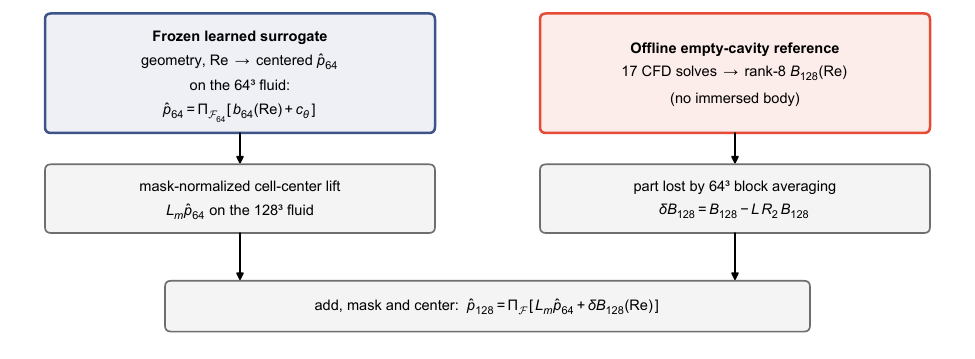}
\caption{\textbf{Reference-corrected lift.} Box outlines: blue, learned; red, CFD-derived and body-free; grey, deterministic operators, including the final assembly (bottom). The lifted surrogate omits the right branch. The no-learning control sets $c_\theta=0$ inside $\hat p_{64}$, keeping the coarse background $b_{64}$. No query pressure and no query-time CFD enter the prediction.}
\label{fig:method}
\end{figure}

\Cref{fig:method} summarizes the predictor.

\paragraph{Frozen coarse surrogate.} The learned component is a $64^3$ surface-conditioned network with 3.15M parameters, developed and frozen before this study. \Cref{app:implementation} specifies it fully; its individual modules are not the subject of this paper.
\begin{itemize}
\item \emph{Output.} It predicts a centered correction $c_\theta$ to a Reynolds-only background $b_{64}(\Re)$, the family-balanced mean of training pressures, and outputs $\hat p_{64}=\Pi_{\mathcal F_{64}}[b_{64}(\Re)+c_\theta]$.
\item \emph{Architecture.} The volume pathway is a masked U-Net hierarchy $64\!\to\!32\!\to\!16\!\to\!8\!\to\!4$ \citep{unet,unet3d}. Nine encoder and decoder stages exchange features with a component-safe graph on the immersed surface and with an area-weighted global surface token.
\item \emph{Inputs.} Six dense channels: the SDF, three coordinates, the outer-wall distance and a scaled $\log\Re$, all zeroed in solid. No velocity, geometry identifier or pressure enters.
\item \emph{Training.} 40{,}500 single-case AdamW updates on the 1{,}620 training cases, with a loss on the complete centered field, the near band, object peaks and gradients. The terminal checkpoint was fixed in advance.
\end{itemize}

\paragraph{Reference correction.} Let $B_{128}(\Re)$ be a centered empty-cavity pressure. It is reconstructed from a rank-8 SVD of 17 native-grid empty-cavity solutions at $\Re\in[10,997]$, with modal coefficients interpolated piecewise-cubically in $\log_{10}\Re$ (\cref{app:reference}). Let $R_2$ be $2\times2\times2$ block averaging and $L$ the corresponding lift. The correction
\begin{equation}\label{eq:subgrid}
\delta B_{128}(\Re)=B_{128}(\Re)-L\,R_2\,B_{128}(\Re)
\end{equation}
is the restriction--prolongation residual of the reference: the part of the empty-cavity field that block averaging followed by the lift discards. It is not an optimal representation residual: an attained numerical least-squares fit of $B_{128}$ by $L$ applied to a $64^3$ field, plus a constant, leaves 17.1\% of its norm at $\Re=10$ and 8.4\% at $\Re=900$, against 21.2\% and 10.6\% for $\delta B_{128}$ (\cref{app:floor}). The predictor is
\begin{equation}\label{eq:predictor}
\hat p_{128}(g,\Re)=\PiF\big[L_m\,\hat p_{64}(g,\Re)+\delta B_{128}(\Re)\big].
\end{equation}
Here $L_m$ is a fluid-mask-normalized trilinear transfer that avoids mixing solid fill into fluid values (\cref{app:metric}). The query geometry $g$ supplies both the $64^3$ and the $128^3$ fluid masks. The predictor accepts only the five training Reynolds numbers, because $b_{64}$ is a lookup table. At its 17 fit points the rank-8 reference reproduces the library fields to a relative error of at most $3.5\times10^{-4}$ ($5.5\times10^{-6}$ at $\Re=10$, $4.8\times10^{-5}$ at $\Re=100$). For $\Re=30,300,900$ it is interpolated; on 16 held-out Reynolds numbers between 11 and 864 the reference reproduces native empty-cavity CFD with mean relative error 0.072\% and worst 0.198\% (\cref{tab:reference}).

\paragraph{Arms.} Four compositions differ only in whether the learned correction $c_\theta$ and the reference correction $\delta B_{128}$ are present:
\begin{itemize}
\item the \emph{lifted surrogate}, $\PiF L_m\hat p_{64}$, which omits the reference;
\item \emph{ours}, \cref{eq:predictor};
\item the \emph{no-learning control}, $\PiF[L_m\Pi_{\mathcal F_{64}}b_{64}+\delta B_{128}]$, which keeps every deterministic term;
\item \emph{background only}, $\PiF L_m\Pi_{\mathcal F_{64}}b_{64}$.
\end{itemize}
Because centering on a fixed mask is linear,
\begin{equation}\label{eq:intervention}
\hat p_{128}-\hat p^{\,\mathrm{lift}}_{128}=\PiF\,\delta B_{128}(\Re),
\end{equation}
so the prospective contrast isolates exactly the reference correction.

\begin{table}[!htb]
\centering\small
\caption{\textbf{Held-out-Reynolds fidelity of the empty-cavity reference} (obstacle-free; 16 held-out values). Relative errors in \%; $1-\rho$ in units of $10^{-6}$ ($\rho$: correlation between the reference and the held-out CFD field). The last column splits the 16 values into low, middle and high Reynolds numbers (5/6/5 values).}
\label{tab:reference}
\begin{tabular}{@{}lrrrr@{}}
\toprule
Held-out Re (16 values, 11--864) & Mean & p90 & Worst & Re 11--37 / 48--205 / 274--864 \\
\midrule
$E_{\mathrm{vol}}$ of $B_{128}(\mathrm{Re})$ (\%) & 0.072 & 0.164 & 0.198 & 0.039 / 0.051 / 0.131 \\
$E_{\nabla}$ of $B_{128}(\mathrm{Re})$ (\%) & 0.071 & 0.162 & 0.244 & -- \\
$1-\rho$, field correlation ($10^{-6}$) & 0.37 & 1.34 & 1.87 & -- \\
\bottomrule
\end{tabular}

\end{table}

\section{Results}\label{sec:results}

\FloatBarrier
\subsection{Prospective effect of the reference}\label{sec:prospective}

\begin{table}[!htb]
\centering\small
\caption{\textbf{Adding the body-free reference to the frozen surrogate} (cohort-mean relative errors, \%). Cohort A is \status{prospective}: both arms were sealed before truth. In cohort B our predictor was sealed, and the lifted arm is its deterministic reconstruction (\status{post hoc}; storage-level agreement checked on five cohort-A cases). Error changes are ratios of cohort means; recall changes are percentage-point differences. For errors, the 95\% $t$ intervals concern the geometric-mean ratio over the 12 geometries; for recall they concern the mean paired geometry-level difference. ``Improved'' counts geometries with a lower (recall: higher) mean. Bold marks cohort-A metrics whose reduction exceeds the predeclared 10\% requirement; the requirement applied to cohort A only.}
\label{tab:main}
\setlength{\tabcolsep}{4.5pt}
\begin{tabular}{@{}l*{6}{r}@{}}
\toprule
 & $E_{\mathrm{vol}}\!\downarrow$ & $E_{\mathrm{near}}\!\downarrow$ & $E_{\nabla}\!\downarrow$ & $E_{\mathrm{pk}}\!\downarrow$ & $E_{\mathrm{obj}}\!\downarrow$ & $R_{\mathrm{obj}}\!\uparrow$ \\
\midrule
\multicolumn{7}{@{}l}{\textit{Cohort A (primary): 60 cases, 12 unseen geometries}}\\
Lifted surrogate & 20.6 & 10.1 & 53.5 & 21.8 & 8.4 & 86.0 \\
Reference-corrected (ours) & \textbf{7.5} & 9.6 & \textbf{12.5} & \textbf{7.2} & 8.4 & 86.0 \\
\quad Change & $-$63.5\% & $-$5.5\% & $-$76.6\% & $-$66.9\% & $+$0.1\% & $+$0.01\,pp \\
\quad 95\% CI (geometry) & {\scriptsize [$-$64.4, $-$62.6]} & {\scriptsize [$-$15.8, $+$6.6]} & {\scriptsize [$-$76.8, $-$76.3]} & {\scriptsize [$-$67.2, $-$66.6]} & {\scriptsize [$-$0.6, $+$1.1]} & {\scriptsize [$-$0.02, $+$0.03]} \\
\quad Geometries improved & 12/12 & 7/12 & 12/12 & 12/12 & 8/12 & 8/12 \\
\midrule
\multicolumn{7}{@{}l}{\textit{Cohort B (second): 60 cases, 12 unseen geometries}}\\
Lifted surrogate & 20.6 & 9.8 & 53.5 & 21.9 & 9.8 & 83.3 \\
Reference-corrected (ours) & 7.4 & 9.8 & 12.5 & 7.3 & 9.7 & 83.3 \\
\quad Change & $-$64.2\% & $-$0.1\% & $-$76.7\% & $-$66.8\% & $-$0.1\% & $+$0.01\,pp \\
\quad 95\% CI (geometry) & {\scriptsize [$-$64.7, $-$63.6]} & {\scriptsize [$-$0.3, 0.0]} & {\scriptsize [$-$76.8, $-$76.5]} & {\scriptsize [$-$67.0, $-$66.6]} & {\scriptsize [$-$0.2, 0.0]} & {\scriptsize [$-$0.01, $+$0.03]} \\
\quad Geometries improved & 12/12 & 10/12 & 12/12 & 12/12 & 10/12 & 7/12 \\
\bottomrule
\end{tabular}

\end{table}

\begin{figure}[!htb]
\centering
\includegraphics{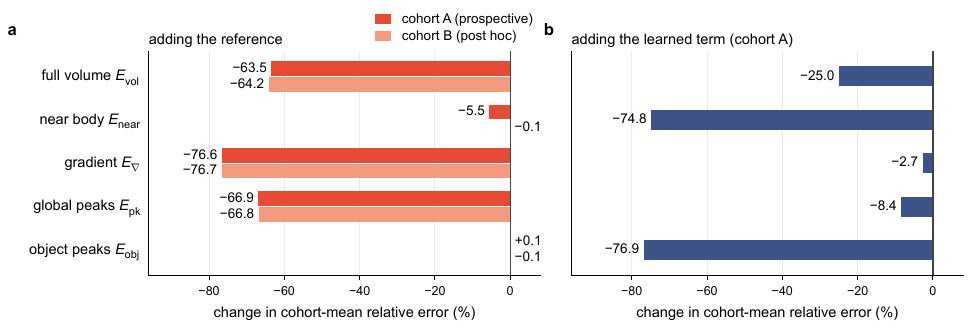}
\caption{\textbf{Metric-level effect of each component}: signed change of the cohort-mean relative error, negative is better, as in \cref{tab:main}. \textbf{a}, Adding the reference to the frozen surrogate (lifted surrogate $\rightarrow$ ours; cohort A \status{prospective}, cohort B \status{post hoc}). Bars are ratios of arithmetic cohort means; no confidence intervals are attached to them. \Cref{tab:main} separately reports intervals for geometry-level geometric-mean ratios, a different estimand. Each cohort has 12 independent geometry groups; capsule-02 dominates the cohort-A near-body mean change, and near-body non-inferiority is not established. \textbf{b}, Adding the learned correction (no-learning control $\rightarrow$ ours; cohort A, \status{retrospective}; cohort means only). Peak-region recall changes by $+0.01$ points in (a) and by $+24.1$ points in (b).}
\label{fig:effects}
\end{figure}

\Cref{tab:main} and \cref{fig:effects}a give the result.
\begin{itemize}
\item \emph{Global metrics.} Adding the reference lowers the global error from 20.6\% to 7.5\% ($-63.5\%$), the gradient error from 53.5\% to 12.5\% ($-76.6\%$) and the global peak error from 21.8\% to 7.2\% ($-66.9\%$). All three decrease in every one of the 12 geometries (enumerated sign-flip $p=0.0005$, the minimum attainable under the sign-symmetry null). The geometry-level reductions span 60.6--65.5\%, 75.4--76.9\% and 66.1--67.6\%. The effect shrinks with Reynolds number: $\Evol$ falls by 67\% at $\Re=10$ but by 50\% at $\Re=900$ (\cref{tab:by_re}).
\item \emph{Object-local metrics.} They do not move: $\Eobj$ changes by $+0.08\%$ and $\Robj$ is unchanged (less than one cell of the top-5\% set on average). Their one-sided 95\% bounds ($+0.95\%$ and $-0.014$ points) lie inside the predeclared tolerances. For $\Enear$ the bound is $+4.28\%$, above the 2\% tolerance, because the capsule-02 outlier widens the interval (\cref{tab:stats}); we therefore do not claim non-inferiority for $\Enear$ in cohort A.
\item \emph{Decision rule.} It is met in 11 of 12 geometries, all three families and all five Reynolds numbers. The exception is capsule-02: its near-body error drops by 47\% while its object-peak error rises by 4.6\%. This single geometry produces the $-5.5\%$ mean change of $\Enear$; the median geometry change is $-0.08\%$ (\cref{sec:boundary}).
\end{itemize}

The diagnostic comparison in \cref{tab:baselines} sharpens the primary result: the body-free reference alone scores 8.9\% globally but 43.9\% near the bodies, whereas the reference-corrected predictor scores 7.5\% and 9.6\%. Rebasing lowers the global score further to 3.2\% without a comparable local gain (\cref{tab:regional}). Reference alone and rebase are \status{retrospective}, not additional prospective tests; neither is the matched no-learning arm.

\begin{table}[!htb]
\centering\small
\caption{\textbf{Diagnostic baselines} (cohort-mean relative errors and recall, \%). Reference alone and rebase are \status{retrospective}; the original cohort-A comparison remains \status{prospective}, and the cohort-B lifted contrast is \status{post hoc} (\cref{tab:main}). The rebased surrogate is \cref{eq:rebase}; the reference alone is the body-free $\PiF B_{128}(\Re)$. $E_{\mathcal J}$ and $E_{\mathcal I}$ are the junction and interior errors of \cref{tab:regional}; $E_{\mathrm{face}}$ is the relative error of differences across interior faces between fluid cells.}
\label{tab:baselines}
\setlength{\tabcolsep}{4pt}
\begin{tabular}{@{}lrrrrrrrrr@{}}
\toprule
Arm & $E_{\mathrm{vol}}\!\downarrow$ & $E_{\mathrm{near}}\!\downarrow$ & $E_{\nabla}\!\downarrow$ & $E_{\mathrm{pk}}\!\downarrow$ & $E_{\mathrm{obj}}\!\downarrow$ & $R_{\mathrm{obj}}\!\uparrow$ & $E_{\mathcal J}$ & $E_{\mathcal I}$ & $E_{\mathrm{face}}$ \\
\midrule
\multicolumn{10}{@{}l}{\emph{Cohort A}} \\
Lifted surrogate & 20.6 & 10.1 & 53.5 & 21.8 & 8.4 & 86.0 & 24.4 & 6.9 & 73.1 \\
Reference-corrected (ours) & 7.5 & 9.6 & 12.5 & 7.2 & 8.4 & 86.0 & 7.8 & 6.9 & 10.1 \\
Reference alone, $\PiF B_{128}$ & 8.9 & 43.9 & 3.6 & 3.3 & 39.3 & 55.8 & 1.4 & 20.5 & 3.3 \\
Rebased on the native reference & 3.2 & 9.4 & 2.5 & 1.2 & 8.2 & 86.0 & 0.7 & 6.7 & 2.6 \\
\multicolumn{10}{@{}l}{\emph{Cohort B}} \\
Lifted surrogate & 20.6 & 9.8 & 53.5 & 21.9 & 9.8 & 83.3 & 24.4 & 6.2 & 73.1 \\
Reference-corrected (ours) & 7.4 & 9.8 & 12.5 & 7.3 & 9.7 & 83.3 & 7.9 & 6.2 & 10.0 \\
Reference alone, $\PiF B_{128}$ & 7.2 & 39.8 & 3.2 & 2.4 & 38.7 & 52.1 & 1.2 & 16.8 & 2.9 \\
Rebased on the native reference & 2.8 & 9.7 & 2.3 & 1.2 & 9.6 & 83.3 & 0.7 & 5.9 & 2.4 \\
\bottomrule
\end{tabular}

\end{table}

\begin{table}[!htb]
\centering\small
\caption{\textbf{Reynolds-number groups} (12 geometries each; the same geometries recur across rows). Errors and recall are cohort means in \%; ``change'' is the relative change of $\Evol$ when adding the reference; ``worst'' is the largest error or smallest recall among the 12 cases.}
\label{tab:by_re}
\begin{tabular}{@{}lrrrrrrrrr@{}}
\toprule
 & & \multicolumn{3}{c}{$E_{\mathrm{vol}}$ (\%)} & \multicolumn{5}{c}{Ours (\%)} \\
\cmidrule(lr){3-5}\cmidrule(l){6-10}
Cohort & Re & Lifted & Ours & Change & $E_{\mathrm{near}}$ & $E_{\mathrm{obj}}$ & $R_{\mathrm{obj}}$ & Worst $E_{\mathrm{obj}}$ & Worst $R_{\mathrm{obj}}$ \\
\midrule
A & 10 & 24.8 & 8.2 & $-$66.8 & 8.8 & 8.6 & 92.2 & 17.1 & 86.8 \\
A & 30 & 24.6 & 8.2 & $-$66.9 & 8.7 & 8.2 & 91.9 & 14.1 & 85.7 \\
A & 100 & 22.8 & 7.6 & $-$66.7 & 9.6 & 8.9 & 88.0 & 15.9 & 75.0 \\
A & 300 & 17.4 & 6.9 & $-$60.5 & 10.7 & 7.4 & 87.8 & 13.5 & 81.0 \\
A & 900 & 13.5 & 6.8 & $-$49.5 & 10.0 & 9.0 & 69.9 & 22.8 & 41.5 \\
\midrule
B & 10 & 25.0 & 8.2 & $-$67.0 & 10.6 & 11.8 & 90.3 & 19.7 & 84.8 \\
B & 30 & 24.8 & 8.2 & $-$67.0 & 10.2 & 11.2 & 89.0 & 19.5 & 80.7 \\
B & 100 & 22.9 & 7.5 & $-$67.2 & 10.7 & 11.1 & 86.3 & 16.6 & 69.8 \\
B & 300 & 17.1 & 6.3 & $-$63.1 & 9.8 & 7.6 & 84.4 & 11.4 & 65.1 \\
B & 900 & 13.4 & 6.7 & $-$49.7 & 7.6 & 7.0 & 66.3 & 19.0 & 43.6 \\
\bottomrule
\end{tabular}

\end{table}

In cohort B, the global, gradient and peak reductions are $-64.2\%$, $-76.7\%$ and $-66.8\%$, with all 12 geometries improved. $\Enear$ and $\Eobj$ change by less than 0.5\% and $\Robj$ by less than 0.1 points in every geometry. The prospective absolute scores of our predictor on cohort B are within 0.3 percentage points of cohort A for the global, gradient, peak and near-body metrics. Object-local scores are somewhat worse ($\Eobj$ 9.7\% vs.\ 8.4\%, $\Robj$ 83.3\% vs.\ 86.0\%).

\subsection{Where the error lives}\label{sec:where}

\Cref{tab:regional} decomposes each arm's error on the region partition of \cref{sec:regions}. The partition reproduces every case's $\Evol$ through \cref{eq:decomposition} to $8\times10^{-16}$.
\begin{itemize}
\item \emph{Junction.} It is 1.5\% of the fluid but holds 68.6\% of the centered truth energy, and 95\% of the lifted surrogate's squared error. The share is nearly independent of Reynolds number (66.6--69.8\%). Adding the reference lowers the junction error from 24.4\% to 7.8\% ($-67.8\%$; interval $[-68.1, -67.5]\%$; 12 of 12 geometries).
\item \emph{Interior.} The interior error changes by $+0.16\%$ (interval $[-0.14, +0.35]\%$). Only 3 of 12 geometries improve, but the interval includes zero; the data do not establish a consistent direction. The reference has a small interior tail.
\item \emph{Cohort B.} The junction change is $-67.7\%$ and the interior change $+0.16\%$.
\item \emph{Gradient and peak metrics.} They are junction effects too (\cref{tab:h10}). The wall shell carries 98.7\% of the truth's gradient energy, and interior $\Egrad$ changes by $+0.09\%$. The 17\% of global peak cells that lie in $\mathcal J$ carry 98.9\% of the lifted surrogate's peak error.
\item \emph{A blind spot of $\Egrad$.} Central differences over two cells do not see the grid-scale checkerboard mode. For the lifted surrogate, the residual after block averaging and trilinear re-extension has 55\% of the original error energy (2.6\% for ours). This operator-residual ratio is not a frequency-band or checkerboard energy fraction: the transfer is not an orthogonal spectral projection. The relative error of face differences between neighbouring fluid cells, which sees this mode, is 73.1\% for the lifted surrogate and 10.1\% for ours (\cref{tab:baselines}). The ranking does not change, but $\Egrad$ understates the error of lifted fields.
\end{itemize}

\begin{table}[!htb]
\centering\small
\caption{\textbf{Regional decomposition} (cohort A unless marked; \status{post hoc}). Cell and truth-energy shares are case means. $E_{\mathcal R}$ applies \cref{eq:region} on region $\mathcal R$ with the canonical full-fluid centering. The last three columns give the change of the cohort-mean regional error: when adding the reference (lifted surrogate $\rightarrow$ ours), when adding the learned term (no-learning control $\rightarrow$ ours; \status{retrospective}), and when adding the reference on cohort B. ``Near body'' is $\mathcal N\cap\mathcal I$; ``--'', not computed for cohort B.}
\label{tab:regional}
\begin{tabular}{@{}lrrrrrrrr@{}}
\toprule
 & \multicolumn{2}{c}{Share (\%)} & \multicolumn{3}{c}{Regional error $E_{\mathcal R}$ (\%)} & \multicolumn{3}{c}{Change of $E_{\mathcal R}$ (\%)} \\
\cmidrule(lr){2-3}\cmidrule(lr){4-6}\cmidrule(l){7-9}
 & & Truth & & & & \multicolumn{2}{c}{Cohort A} & Cohort B \\
Region $\mathcal R$ & Cells & energy & Lifted & No-learning & Ours & +Reference & +Learned & +Reference \\
\midrule
Junction $\mathcal J$ & 1.5 & 68.6 & 24.4 & 7.9 & 7.8 & $-$67.8 & $-$0.3 & $-$67.7 \\
Rest of wall shell $\mathcal S\setminus\mathcal J$ & 31.7 & 18.0 & 5.7 & 9.1 & 5.3 & $-$6.7 & $-$42.2 & -- \\
Interior $\mathcal I$ & 66.9 & 13.4 & 6.9 & 17.3 & 6.9 & $+$0.2 & $-$59.9 & $+$0.2 \\
\quad Near body & 12.3 & 2.1 & 10.0 & 40.0 & 10.0 & 0.0 & $-$75.0 & -- \\
\quad Elsewhere & 54.6 & 11.4 & 6.3 & 10.7 & 6.3 & $+$0.3 & $-$40.6 & -- \\
\bottomrule
\end{tabular}

\end{table}

\begin{table}[!htb]
\centering\small
\caption{\textbf{Gradient and global-peak errors by region} (cohort A, \status{post hoc}). $^\ast$Share of the global top-5\% peak support rather than of energy. The last column is the median over cases of our regional error divided by the lifted one; for $\Epk$ the $\mathcal J$ mask is the audit's lid-edge band, whose per-case errors coincide with those on $\mathcal J$.}
\label{tab:h10}
\setlength{\tabcolsep}{4pt}
\begin{tabular}{@{}llrrrrrr@{}}
\toprule
 & & Truth & \multicolumn{2}{c}{Share of sq.\ error (\%)} & \multicolumn{2}{c}{Regional error (\%)} & Median ratio \\
\cmidrule(lr){4-5}\cmidrule(lr){6-7}
Metric & Region & share (\%) & Lifted & Ours & Lifted & Ours & ours/lifted \\
\midrule
$E_{\nabla}$ & Wall shell $\mathcal S$ & 98.7 & 99.7 & -- & 53.8 & 12.3 & -- \\
 & Junction $\mathcal J$ & 94.5 & 99.4 & 90.1 & -- & -- & -- \\
 & Interior $\mathcal I$ & 1.3 & 0.3 & 5.3 & 22.6 & 22.6 & -- \\
\midrule
$E_{\mathrm{pk}}$ & Junction $\mathcal J$ & 17.2$^\ast$ & 98.9 & 92.5 & -- & -- & 0.32 \\
 & Interior $\mathcal I$ & 32.1$^\ast$ & 0.3 & 2.2 & -- & -- & 1.01 \\
\bottomrule
\end{tabular}

\end{table}

\begin{figure}[!htb]
\centering
\includegraphics{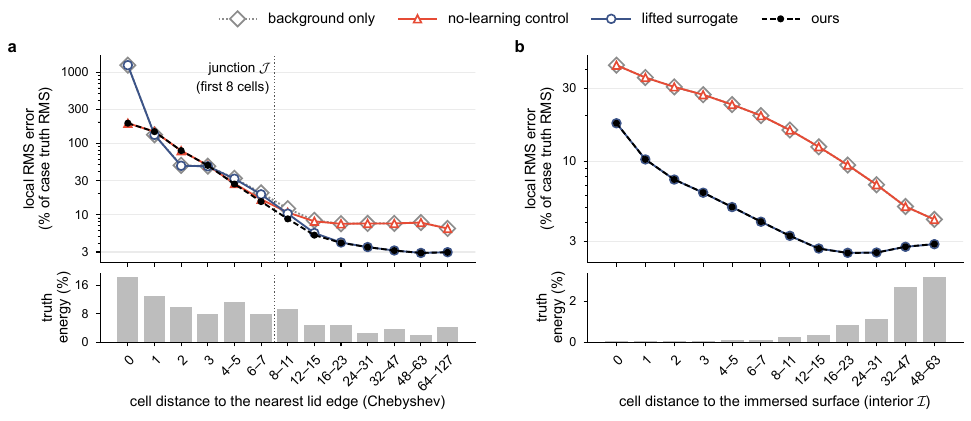}
\caption{\textbf{Error profiles} (cohort A, 60 cases; \status{post hoc}; the two arms without the learned term are \status{retrospective}). Local RMS error per distance bin as a percentage of the case's full-fluid truth RMS, averaged over cases (log scale). Bins widen with distance but are drawn at equal spacing. The upper curves are per-bin RMS values, not quantities that accumulate linearly with the number of cells; their values can still change with bin boundaries and widths; the lower strips give each bin's share of truth energy, which grows with the number of cells in the bin. ``Background only'' is the lifted Reynolds-only background (\cref{sec:method}), with neither the learned correction nor the reference; away from the lid edges it coincides with the no-learning control. \textbf{a}, Chebyshev cell distance to the nearest lid edge; the dotted line bounds the 8-cell junction. Near the edges our predictor (black) follows the no-learning control (red); far from them it follows the lifted surrogate (blue). Markers follow \cref{fig:thesis}c: open circle, lifted; triangle, no-learning control; filled dot, ours. \textbf{b}, Cell distance to the immersed surface within the interior $\mathcal I$. The lifted and reference-corrected curves agree to within 0.2\% (the blue curve lies under the black one), and the arms without the learned term are 1.4--4.9$\times$ worse.}
\label{fig:profiles}
\end{figure}

\paragraph{Anatomy of the junction.} \Cref{tab:junction} splits $\mathcal J$ as the physics suggests. The two edge blocks perpendicular to the lid motion, where the lid scrapes against a side wall, hold 59.9\% of the truth energy in 0.69\% of the fluid; the four vertex blocks add 7.9\%. The two edges parallel to the motion hold only 0.8\%, consistent with a bounded pressure there. \Cref{fig:profiles}a resolves the junction by distance to the nearest lid edge. The single layer of cells on the edges holds 18.5\% of the truth energy and, on average over cases, 88\% of the lifted surrogate's squared error. There the lifted local RMS error is about $12.5\times$ the case's truth RMS, and ours is about $1.9\times$. In the next two cell layers the reference over-corrects slightly; from 16 cells onward the lifted and reference-corrected curves agree to within about 1\%. In the interior (\cref{fig:profiles}b) the reference changes nothing at any distance from the body, whereas removing the learned term raises the local error by a factor of 1.4 to 4.9.

\begin{table}[!htb]
\centering\small
\caption{\textbf{Junction anatomy} (cohort A, \status{post hoc}; case means; cells are counts, all other values in \%). Blocks along the two edges perpendicular to the lid motion (excluding the vertex blocks), along the two parallel edges, and the four $8^3$ vertex blocks. The last column evaluates the body-free reference $\PiF B_{128}(\Re)$ alone against the truth.}
\label{tab:junction}
\begin{tabular}{@{}lrrrrrr@{}}
\toprule
 & & Truth energy & \multicolumn{4}{c}{Regional error (\%)} \\
\cmidrule(l){4-7}
Subregion & Cells & share (\%) & Lifted & No-learning & Ours & $\Pi_{\mathcal F}B_{128}$ alone \\
\midrule
Perpendicular edges ($x{=}0,1$) & 14,336 & 59.9 & 24.4 & 7.7 & 7.7 & 1.1 \\
Parallel edges ($z{=}0,1$) & 14,336 & 0.8 & 5.7 & 7.5 & 5.4 & 8.1 \\
Vertex blocks & 2,048 & 7.9 & 25.4 & 8.9 & 8.9 & 1.2 \\
Junction $\mathcal J$ (all) & 30,720 & 68.6 & 24.4 & 7.9 & 7.8 & 1.4 \\
\bottomrule
\end{tabular}

\end{table}

\begin{table}[!htb]
\centering\small
\caption{\textbf{Sensitivity to the junction width} (cohort A, \status{post hoc}). $\mathcal J_k$ contains the cells within $k$ cells of a lid edge (the bins of \cref{fig:profiles}a); changes are those of the cohort-mean regional error when adding the reference. ``Outside'' includes the rest of the wall shell; for $k\le2$ it contains the cell layers in which the reference over-corrects slightly.}
\label{tab:width}
\begin{tabular}{@{}rrrrrr@{}}
\toprule
Width $k$ & Cells & Truth energy & Lifted sq.\ error & Change inside & Change outside \\
(cells) & (\%) & share (\%) & share (\%) & $\mathcal J_k$ (\%) & $\mathcal J_k$ (\%) \\
\midrule
1 & 0.02 & 18.5 & 88.4 & $-$85.0 & $+$6.8 \\
2 & 0.10 & 31.5 & 91.3 & $-$75.3 & $+$3.9 \\
4 & 0.38 & 49.3 & 93.2 & $-$69.7 & $-$9.0 \\
8 & 1.47 & 68.6 & 95.1 & $-$67.8 & $-$3.3 \\
12 & 3.20 & 77.9 & 95.7 & $-$67.3 & $-$0.8 \\
16 & 5.49 & 82.7 & 96.0 & $-$67.0 & $-$0.3 \\
\bottomrule
\end{tabular}

\end{table}
\needspace{5\baselineskip}\paragraph{What the reference corrects.} Three observations identify its target.
\begin{itemize}
\item \emph{It does not depend on the body.} The body-free field $\PiF B_{128}$ alone matches the truth in $\mathcal J$ to 1.4\%, and in the perpendicular-edge blocks to 1.1\% (\cref{tab:junction}). This holds at every Reynolds number (1.2--1.5\%) and stays below 3.7\% in every case; the largest geometry mean, 2.8\%, belongs to capsule-02, whose near-body band reaches into the junction.
\item \emph{Its settings match the truth.} The reference library and the test truth use byte-identical discretization and solver settings (\cref{app:cfd}).
\item \emph{The estimated representation residual is concentrated at the lid edges.} A numerical full-fluid least-squares fit of $L_m$ applied to a $64^3$ field leaves 13.8\% of the truth norm on average. That global fit's regional residuals are 16.6\% in $\mathcal J$, 2.9\% in the interior near-body band and 1.0\% in the interior; they are not independently optimized regional lower bounds. Of the fitted residual energy, 99.9\% lies in $\mathcal J$ (\cref{app:floor}).
\end{itemize}

The reduction in $\Evol$ is therefore agreement with the solver's own lid-edge values, not an improvement of the pressure response to the body. The 69\% junction share is also a property of this grid rather than of the continuum. Near a $1/r$ singularity the energy grows logarithmically with refinement, and in the empty cavity the share of a band of fixed physical width grows from $64^3$ to $256^3$ while that of a band of fixed cell count falls (\cref{sec:physics}). The correction removes about two thirds of the junction error. Yet the junction still holds 75\% of our predictor's remaining squared error, because the lifted coarse field, not the reference, sets the error there: 7.8\% for the reference-corrected predictor versus 1.4\% for the reference alone. \Cref{sec:rebase} tests the obvious consequence.

\paragraph{A frozen audit label.} Before this regional analysis we froze a separate audit with a descriptive rule. It labels the reference gain ``boundary dominated'' only if every one of its inputs is below a small limit. That rule returns ``spatially mixed''. Its only failing input is the near-body squared-error change pooled over geometries at each Reynolds number ($-22\%$ to $-61\%$), and capsule-02's junction cells dominate it. Every other input is below $10^{-4}$ against a limit of $0.05$. Excluding capsule-02, as a post hoc sensitivity analysis, the pooled near-body changes lie within $\pm0.5\%$ at every Reynolds number. We report the frozen label unchanged (\cref{app:more}).

\FloatBarrier
\subsection{Diagnostic baselines: the reference alone and a rebased surrogate}\label{sec:rebase}
The correction $\delta B_{128}$ adds only what block averaging removes from $B_{128}$. The surrogate's own coarse representation of the lid edges, learned from $64^3$ training data, stays in the prediction. A \emph{rebased} surrogate replaces it by the native reference:
\begin{equation}\label{eq:rebase}
\hat p^{\,\mathrm{rebase}}_{128}=\PiF\big[B_{128}+L_m(\hat p_{64}-B_{64})\big]=\hat p^{\,\mathrm{lift}}_{128}+\PiF\big[B_{128}-L_mB_{64}\big].
\end{equation}
Here $B_{64}(\Re)$ is an empty-cavity CFD solution on the $64^3$ grid with the settings of the reference library. Like $B_{128}$ it is body-free and computed offline, once per Reynolds number. The rebased surrogate therefore differs from ours only in its body-independent term. We also evaluate the reference alone, $\PiF B_{128}$, which contains no learned body-conditioned response; it still uses the query fluid mask for centering and evaluation. Both are \status{retrospective} diagnostics, designed after both cohorts had been evaluated; cohort B uses the reconstructed lifted arm.

\Cref{tab:baselines} gives three results, each replicated on cohort B.
\begin{itemize}
\item \emph{Similar global scores can hide different body accuracy.} The reference alone has a global error of 8.9\% (A) and 7.2\% (B), close to ours (7.5\% and 7.4\%), while its near-body error is 44\% and 40\%. It also beats the body-aware lifted model on all three global metrics in cohort A (volume: 8.9\% versus 20.6\%; gradient: 3.6\% versus 53.5\%; global peaks: 3.3\% versus 21.8\%), despite its much worse near-body error (43.9\% versus 10.1\%). These are retrospective diagnostic comparisons, not an additional prospective superiority test.
\item \emph{Rebasing lowers every global metric further.} $\Evol$ falls from 7.5\% to 3.2\% (A) and from 7.4\% to 2.8\% (B), and the junction error from 7.8\% to 0.7\%. The per-geometry ratio of $\Evol$ to ours is 0.29--0.60 (A) and 0.31--0.46 (B).
\item \emph{Cohort-mean body metrics change little.} $\Enear$, $\Eobj$, $\Robj$ and the interior error move by at most 0.33 points, and the interior near-body error by at most 0.1 points. The largest per-geometry change of $\Enear$, $-30\%$, is again capsule-02.
\end{itemize}
Our composition is thus not the best way to use the reference, and we do not claim it is. Its role in this paper is that of a controlled intervention: \cref{eq:intervention} isolates one body-independent term, whose effect can be located exactly. The rebased surrogate shows that most of the remaining global error of ours can also be removed by a body-independent term, with little change in the cohort-mean body metrics. Importantly, the junction's share of error is model-dependent: its case-mean $w_{\mathcal J}$ in cohort A is 95.1\% for lifted, 75.4\% for ours, but only 1.9\% for reference alone and 4.0\% for rebase. These post hoc shares are computed per case before averaging, not by multiplying mean truth weights and mean errors. A large truth-energy share therefore does not force a predictor's error to lie in the junction.

\FloatBarrier
\subsection{The learned term carries the body response}\label{sec:learned}

\begin{figure}[!htb]
\centering
\includegraphics{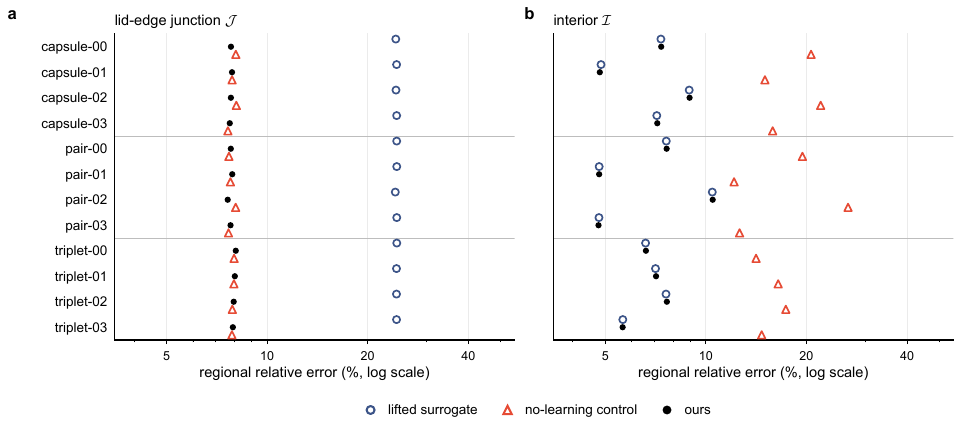}
\caption{\textbf{Separable roles in every geometry} (cohort A; \status{post hoc} regional analysis, \status{retrospective} no-learning control). Per-geometry mean regional error in the lid-edge junction (\textbf{a}) and in the interior (\textbf{b}): lifted surrogate (upper lane), ours (middle) and no-learning control (lower lane). Horizontal rules separate the capsule, pair and triplet families. In every geometry, adding the reference changes the junction error by $-68.5\%$ to $-67.0\%$ and the interior error by $-0.8\%$ to $+0.6\%$; adding the learned term changes the interior error by $-68.0\%$ to $-53.2\%$ and the junction error by $-5.3\%$ to $+1.4\%$.}
\label{fig:geometry}
\end{figure}

The no-learning control keeps the coarse background, both masks, the transfer, the reference correction and the centering, and removes only the learned correction (\cref{tab:controls}; \status{retrospective}). It was constructed without reading any pressure truth, but after cohort A had been inspected.
\begin{itemize}
\item \emph{Metric level.} Relative to the control, our predictor lowers $\Enear$ by 74.8\%, $\Eobj$ by 76.9\% and $\Evol$ by 25.0\%, and it raises $\Robj$ by 24.1 points. $\Enear$ and $\Eobj$ improve in all 12 geometries, in every family and at every Reynolds number.
\item \emph{Region level.} The learned term mirrors the reference (\cref{tab:regional}, \cref{fig:geometry}). It removes 59.9\% of the interior error and 75.0\% of the interior near-body error, but only 0.3\% of the junction error. The separation holds in every geometry: each component moves the other's region by at most 5.3\% (\cref{fig:geometry}).
\item \emph{Body-local is not body-induced.} Near the bodies most of the pressure is still cavity flow. The reference-relative disturbance $d$ of \cref{eq:body-response} carries 23.5\% of the truth energy in the interior near-body band, 19.3\% on the object-peak support, 4.9\% in the interior and 0.03\% in $\mathcal J$ (means of case-level energy ratios). Its smallest energy ratio in the interior near-body band is 0.00582. Measured against $\|d\|$ instead of $\|\PiF p\|$, the interior near-body error is 24.6\% for ours, 24.7\% for the lifted surrogate and 88.8\% for the no-learning control; on the object-peak support it is 26.0\%. In the interior near-body band, the ratio of the two cohort-mean error normalizations is about 2.5; this is not a per-case conversion factor. The correction barely changes this reference-relative body-response diagnostic.
\end{itemize}

\begin{table}[!htb]
\centering\small
\caption{\textbf{Four compositions and two retrospective comparisons} (cohort A, cohort-mean relative errors in \%). \status{p}: prospective; \status{r}: retrospective, an arm or retraining created after the cohort had been inspected. The seed replication and the matched FNO use the same data, inputs, background, 40{,}500 updates, lift, reference and evaluator. The FNO has 96.8\% of the surrogate's real-valued parameters.}
\label{tab:controls}
\begin{tabular}{@{}llrrrrrr@{}}
\toprule
Arm (cohort A) & Status & $E_{\mathrm{vol}}\!\downarrow$ & $E_{\mathrm{near}}\!\downarrow$ & $E_{\nabla}\!\downarrow$ & $E_{\mathrm{pk}}\!\downarrow$ & $E_{\mathrm{obj}}\!\downarrow$ & $R_{\mathrm{obj}}\!\uparrow$ \\
\midrule
Background only & \textsc{r} & 21.8 & 38.3 & 53.6 & 22.1 & 36.4 & 61.9 \\
Lifted surrogate & \textsc{p} & 20.6 & 10.1 & 53.5 & 21.8 & 8.4 & 86.0 \\
No-learning control & \textsc{r} & 10.1 & 38.0 & 12.9 & 7.9 & 36.4 & 61.8 \\
Reference-corrected (ours) & \textsc{p} & 7.5 & 9.6 & 12.5 & 7.2 & 8.4 & 86.0 \\
\midrule
Ours, 3 retrained seeds (mean) & \textsc{r} & 7.4 & 9.6 & 12.5 & 7.2 & 8.5 & 85.6 \\
Matched residual FNO, 3 seeds (mean) & \textsc{r} & 7.9 & 12.6 & 12.7 & 7.3 & 11.4 & 81.7 \\
\bottomrule
\end{tabular}

\end{table}

Denote the four arms' errors by $e_{00}$ (background only), $e_{10}=e_{00}+\ell$ (learned term added), $e_{01}=e_{00}+r$ (reference added) and $e_{11}=e_{00}+\ell+r$. Here $\ell$ is the lifted learned correction and $r=\PiF\delta B_{128}$. Because every operator is linear for fixed masks, the interaction of squared relative errors is
\begin{equation}
(E^2_{00}-E^2_{10})-(E^2_{01}-E^2_{11})=2\langle \ell, r\rangle/\|\PiF p\|^2,
\end{equation}
whose unnormalized numerator, on fixed support, does not depend on the truth. The normalized interaction does depend on the truth energy and, on truth-selected peak regions, on that support. Its case mean is $+9.7\times10^{-6}$, 0.2\% of the smaller main effect (per-case range $-1.9\times10^{-4}$ to $+6.7\times10^{-5}$). The small interaction shows near-additivity of squared losses in these cases, not disjoint spatial support: overlapping zero-mean vectors can have zero inner product. Regional improvements provide separate evidence about where the corrections help. The identity also holds on every region with the regional truth energy in the denominator. There the interaction is at most $2.3\times10^{-3}$ in any case and region, including the object-peak support, and at most $1.2\times10^{-3}$ for capsule-02. The square of $\Eobj$ also obeys this identity because its truth-selected support is fixed across the four arms. Recall $\Robj$ does not: its prediction-selected support varies. Capsule-02's object-peak error can worsen through alignment of the added reference with the existing error, without a change of truth-selected support (\cref{sec:boundary}).

\paragraph{A matched spectral comparator.} Is this local accuracy specific to the learned surrogate? We trained a residual masked FNO \citep{fno} under the same contract (\status{retrospective}). It uses the same 1{,}620 training cases, deployment-visible inputs, Reynolds-only background and 40{,}500 updates, three predeclared seeds, and the same lift, reference and evaluator, so the comparison is between two learned surrogates inside the same composition. The FNO has four spectral blocks, width 21 and $6^3$ retained modes out of 32 per axis, for 3.05M real-valued parameters (96.8\% of the surrogate's). Its optimizer, learning rate and schedule were inherited from the surrogate and not tuned. The mode count was kept from an earlier FNO baseline, and the width was set to match capacity. The FNO reaches similar global accuracy but is worse locally (\cref{tab:controls}). Averaged over seeds, the reference-corrected surrogate has 24.3\% lower $\Enear$, 25.6\% lower $\Eobj$ and 3.9 points higher recall. For every seed it passes all nine checks of a predeclared symmetric superiority rule, and no FNO seed passes the reverse rule (\cref{app:fno}). The strong low-pass restriction of 6 modes is a plausible reason for the local deficit. The result is evidence against one untuned FNO configuration, not against the FNO family.

\paragraph{Seed sensitivity.} We retrained the surrogate from scratch with three predeclared seeds under the same configuration (\status{retrospective}). The cohort means changed by at most 0.08 points ($\Evol$), 0.34 ($\Enear$), 0.23 ($\Eobj$) and 0.37 ($\Robj$). Including the originally sealed run, the four-run ranges are 0.18, 0.34, 0.23 and 0.51 points (\cref{tab:seeds}). The tails vary more: the seed ranges of the 90th percentiles of $\Enear$ and $\Eobj$ are 1.1 and 0.9 points. An unblinded AI-agent visual review of all 24 fixed-scale panels (8 cases $\times$ 3 seeds) reported no new artifacts. The effects of \cref{sec:prospective,sec:learned} are one to two orders of magnitude larger than these ranges; the 5.5\% mean change of $\Enear$ in cohort A is not.

\begin{table}[!htb]
\centering\small
\caption{\textbf{Seed and platform sensitivity} (cohort A, \status{retrospective}; mean relative errors and recall, \%).}
\label{tab:seeds}
\begin{tabular}{@{}l*{6}{r}@{}}
\toprule
Run (cohort A) & $E_{\mathrm{vol}}\!\downarrow$ & $E_{\mathrm{near}}\!\downarrow$ & $E_{\nabla}\!\downarrow$ & $E_{\mathrm{pk}}\!\downarrow$ & $E_{\mathrm{obj}}\!\downarrow$ & $R_{\mathrm{obj}}\!\uparrow$ \\
\midrule
Historical (sealed; Table~\ref{tab:main}) & 7.54 & 9.57 & 12.53 & 7.23 & 8.41 & 85.97 \\
\midrule
Retrained, seed 42 & 7.44 & 9.66 & 12.52 & 7.19 & 8.37 & 85.46 \\
Retrained, seed 314159 & 7.43 & 9.70 & 12.53 & 7.21 & 8.59 & 85.55 \\
Retrained, seed 271828 & 7.36 & 9.35 & 12.52 & 7.20 & 8.40 & 85.83 \\
\midrule
Range, 3 retrainings & 0.08 & 0.34 & 0.01 & 0.02 & 0.23 & 0.37 \\
Range, all 4 runs & 0.18 & 0.34 & 0.02 & 0.04 & 0.23 & 0.51 \\
\bottomrule
\end{tabular}

\end{table}

\FloatBarrier
\subsection{One observed failure case and the remaining errors}\label{sec:boundary}

\paragraph{When the near field reaches the lid edge.} The local-tolerance interpretation has one material exception, not a failure of the mutually exclusive junction/interior partition. The squared-error interaction remains small there (regional interaction at most $1.2\times10^{-3}$), but capsule-02 has by far the largest share of its near-body band in the junction $\mathcal J$ (1.12\%; \cref{tab:geometry}), where the reference acts. The only other geometry with any, cohort-B capsule-03, has 0.42\% and an $\Enear$ change of $-0.5\%$.
\begin{itemize}
\item Its near-body error falls by 47\% when the reference is added. The gain comes entirely from the junction cells: its near-body error outside the wall shell changes by $+0.01\%$, and 92\% of its lifted near-band error lies in the shell.
\item At the same time its object-peak error rises by 4.6\%, the largest local harm in either cohort.
\item Proximity to the lid alone does not produce this effect. Cohort B contains a capsule as close to the lid (gap 0.070 vs.\ 0.078), with 10.4\% of its band inside the lid shell but none in the junction; its $\Enear$ changes by $+0.3\%$.
\end{itemize}

\begin{table}[!htb]
\centering\small
\caption{\textbf{Per-geometry effect of adding the reference} (cohort A, means over five Reynolds numbers). ``Lid gap'' is the solid-to-lid distance in units of the box side; $\mathcal N\cap\mathcal J$ is the share of the near-body band inside the lid-edge junction. Placement strata refer to the generator's target faces; $z{=}1$ is not the lid.}
\label{tab:geometry}
\begin{tabular}{@{}llrrrrrr@{}}
\toprule
 & & Lid gap & $\mathcal N\cap\mathcal J$ & \multicolumn{4}{c}{Change when adding the reference} \\
\cmidrule(l){5-8}
Geometry & Placement & (box side) & (\% of $\mathcal N$) & $E_{\mathrm{vol}}$ (\%) & $E_{\mathrm{near}}$ (\%) & $E_{\mathrm{obj}}$ (\%) & $R_{\mathrm{obj}}$ (pp) \\
\midrule
capsule-00 & Interior & 0.242 & 0.00 & $-$63.3 & $-$0.5 & $-$0.6 & $+$0.04 \\
capsule-01 & Near $z{=}1$ & 0.328 & 0.00 & $-$65.1 & $-$0.3 & $-$0.4 & $+$0.02 \\
capsule-02 & Near $x$/$y$ face & 0.078 & 1.12 & $-$62.2 & $-$47.4 & $+$4.6 & $-$0.01 \\
capsule-03 & Near $z{=}0$ & 0.516 & 0.00 & $-$62.8 & $+$0.2 & $+$0.1 & $+$0.01 \\
pair-00 & Interior & 0.492 & 0.00 & $-$62.7 & $+$0.1 & $-$0.2 & $+$0.01 \\
pair-01 & Near $z{=}1$ & 0.484 & 0.00 & $-$65.3 & 0.0 & $-$0.2 & $+$0.04 \\
pair-02 & Near $x$/$y$ face & 0.250 & 0.00 & $-$60.6 & $-$0.1 & $+$0.1 & $-$0.06 \\
pair-03 & Near $z{=}0$ & 0.539 & 0.00 & $-$65.5 & 0.0 & $-$0.1 & $+$0.02 \\
triplet-00 & Interior & 0.367 & 0.00 & $-$63.1 & $-$0.1 & 0.0 & $-$0.04 \\
triplet-01 & Near $z{=}1$ & 0.273 & 0.00 & $-$63.2 & $-$0.1 & $-$0.2 & $-$0.06 \\
triplet-02 & Near $x$/$y$ face & 0.305 & 0.00 & $-$63.1 & $+$0.1 & $-$0.1 & $+$0.05 \\
triplet-03 & Near $z{=}0$ & 0.266 & 0.00 & $-$64.7 & 0.0 & 0.0 & $+$0.04 \\
\bottomrule
\end{tabular}

\end{table}

Our hypothesis is that the reference affects body metrics only where the body's near field reaches materially into the singular corner region. One case supports it and one post hoc comparison is consistent with it, so it is not a characterization.

\begin{figure}[!htb]
\centering
\includegraphics{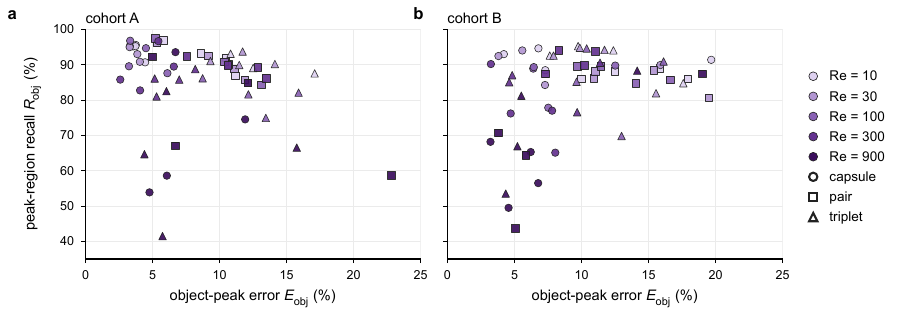}
\caption{\textbf{Object-local accuracy per case} of our predictor (\textbf{a}, cohort A; \textbf{b}, cohort B; both prospective absolute scores): object-peak error against peak-region recall. Colour encodes the Reynolds number (darker is higher) and marker shape the family. Low recall concentrates at $\Re\ge300$, where the full-volume error is lowest (\cref{tab:by_re}).}
\label{fig:tails}
\end{figure}

\begin{figure}[!htb]
\centering
\includegraphics{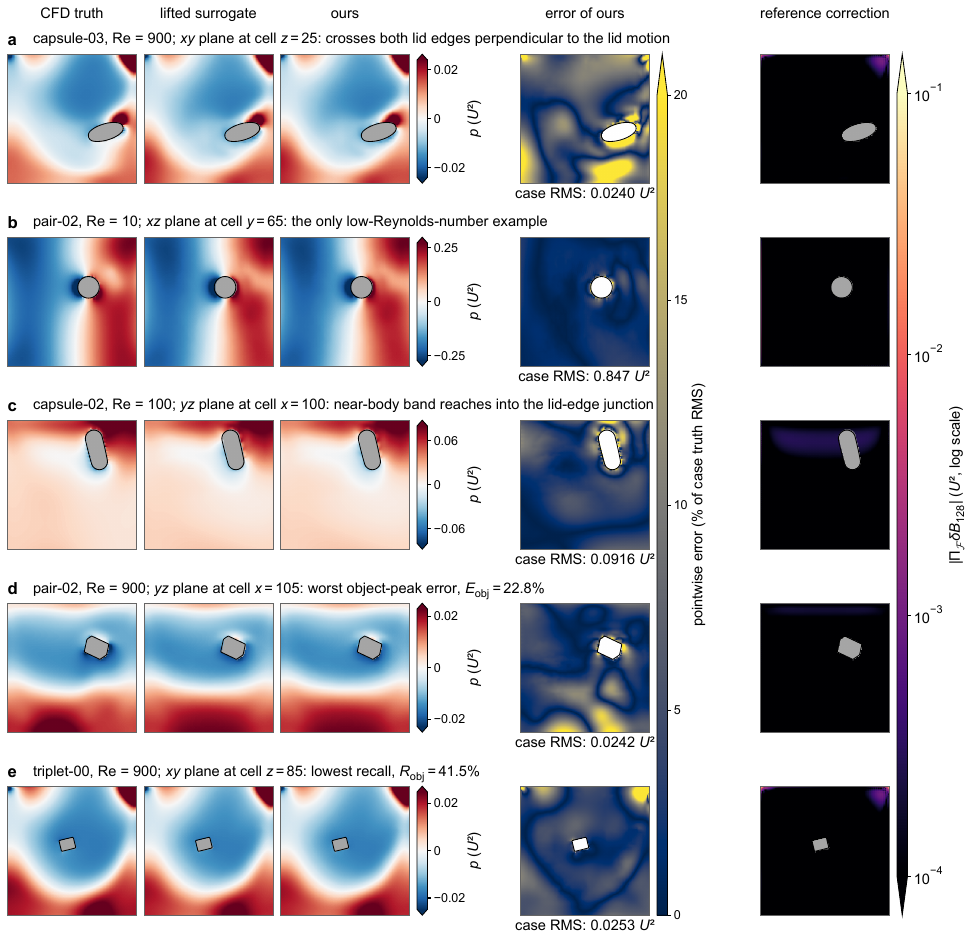}
\caption{\textbf{Fields on native $128^3$ planes} (cohort A; pressure in units of $U^2$). Rows \textbf{a}--\textbf{e}: a plane crossing both perpendicular lid edges; the only low-Reynolds-number example; capsule-02, whose near-body band reaches into the lid-edge junction; the worst object-peak case; the lowest-recall case. The first four cases are those of the original visual protocol. In $xy$ and $yz$ planes the lid is the top edge. Columns 1--3 share a symmetric scale set to the 99th percentile of $|p|$ \emph{outside} the wall shell, so the lid corners saturate by design. Column 4 is the pointwise error of our predictor as a percentage of the case's full-fluid truth RMS, whose RMS over the fluid equals $100\,\Evol$; the case RMS printed below converts it to $U^2$. Its scale, 0--20\%, is the same in every row and in \cref{fig:favorable}, so rows can be compared directly. Column 5 shows $|\hat p_{\mathrm{corrected}}-\hat p_{\mathrm{lifted}}|$ computed from the stored, unclipped fields, corresponding to the magnitude of $\PiF\delta B_{128}$ up to storage precision. Its fixed logarithmic scale spans $10^{-4}$--$10^{-1}\,U^2$; finite fluid values below the lower limit, including exact zeros, use the floor colour. Values above the upper limit saturate. The correction reaches $0.3\,U^2$ at the perpendicular lid edges (top corners of rows a and e); row b has small but nonzero edge corrections, up to $0.0042\,U^2$. Grey is solid in pressure and correction panels and white in error panels; black contours mark the zero level of the signed distance. Small-scale colour variations are visible near the voxelized solid--fluid interfaces.}
\label{fig:fields}
\end{figure}

\paragraph{Object-local tails at high Reynolds number.} The composition does not solve the local pressure structure around bodies. \Cref{fig:tails} shows the case-level object-peak error and recall in both cohorts.
\begin{itemize}
\item At $\Re=900$ mean recall drops to 69.9\% (A) and 66.3\% (B), from over 90\% at $\Re=10$, even though the global error is lowest at $\Re\geq300$.
\item The minimum recall, 41.5\% (A) and 43.6\% (B), occurs in multi-body cases.
\item The worst object-peak errors are 22.8\% (A) and 19.7\% (B).
\end{itemize}
\Cref{fig:fields} shows the worst object-peak and lowest-recall cases next to three reference cases, with object-attached halos and displaced lobes. On 120 training cases, the specified coarse and native target-informed oracles attain near-body errors of 2.65\% and 2.49\%, and object-peak errors of 3.19\% and 2.97\%, respectively (\cref{app:lineage}). This comparison does not quantify the fraction of representable corrections retained at $64^3$. In a separate development-data diagnostic, the tested residual learners did not reliably reproduce the target-informed gain; this does not identify a unique learning or representation bottleneck. \Cref{app:more} gives per-Reynolds, per-family and tail statistics, all case scores and the best cohort-A cases.

\subsection{Numerical robustness: grid, lid and iterative convergence}\label{sec:physics}
Three questions remain after \cref{sec:where}. Is the junction dominance a property of the singular lid, how does it depend on the grid, and does it depend on how tightly the test truth was converged? We answered them with new CFD under protocols frozen before any run (\cref{app:physics,app:h6}; \status{retrospective}, descriptive). The regularized lid uses $u_x=[1-(2x-1)^{18}]^2[1-(2z-1)^{18}]^2$ \citep{leriche2000,bouffanais2007}, which vanishes at the lid edges. Everything else is unchanged.
\begin{itemize}
\item \emph{Grid refinement.} We solved the empty cavity at $\Re=10,100,997$ on $64^3$, $128^3$ and $256^3$ grids for both lids. The $256^3$ runs were initialized from the converged $128^3$ solution, and all runs met the residual criteria.
\item \emph{Regularized cohort A.} We re-solved all 60 cohort-A cases at $128^3$ with the regularized lid. We also refitted the reference, by the unchanged procedure, to 17 regularized empty-cavity solutions. All 77 runs converged.
\end{itemize}

\begin{table}[!htb]
\centering\small
\caption{\textbf{Grid dependence of the empty cavity} (\status{retrospective}; shares of the centered pressure energy, \%). ``8 cells'': share of the cells within eight cells of a lid edge at each resolution. ``Width $1/16$'': share of the band of that physical width, the same region at every resolution and equal to the 8-cell band at $128^3$. ``$\delta B$'': energy of the restriction--prolongation residual $\delta B=(I-LR_2)B$ relative to that of $B$.}
\label{tab:grid}
\setlength{\tabcolsep}{4pt}
\begin{tabular}{@{}llrrrrrrrrr@{}}
\toprule
 & & \multicolumn{3}{c}{8 cells} & \multicolumn{3}{c}{width $1/16$} & \multicolumn{3}{c}{$\delta B$} \\
\cmidrule(lr){3-5}\cmidrule(lr){6-8}\cmidrule(l){9-11}
Lid & $\Re$ & $64^3$ & $128^3$ & $256^3$ & $64^3$ & $128^3$ & $256^3$ & $64^3$ & $128^3$ & $256^3$ \\
\midrule
sharp & 10 & 83.9 & 72.5 & 62.7 & 66.6 & 72.5 & 76.6 & 5.48 & 4.51 & 3.85 \\
sharp & 100 & 85.1 & 71.4 & 58.6 & 66.4 & 71.4 & 74.9 & 4.71 & 3.79 & 3.24 \\
sharp & 997 & 73.1 & 65.0 & 44.8 & 63.0 & 65.0 & 66.1 & 2.72 & 1.11 & 0.68 \\
\addlinespace
regularized & 10 & 74.1 & 48.9 & 24.2 & 47.2 & 48.9 & 49.4 & 0.36 & 0.06 & 0.01 \\
regularized & 100 & 78.0 & 51.4 & 24.1 & 50.0 & 51.4 & 51.8 & 0.30 & 0.04 & 0.01 \\
regularized & 997 & 73.0 & 60.8 & 30.9 & 59.8 & 60.8 & 60.9 & 0.43 & 0.03 & 0.00 \\
\bottomrule
\end{tabular}

\end{table}

\Cref{tab:grid} gives the grid dependence.
\begin{itemize}
\item \emph{Unregularized lid.} The mean pressure magnitude in the lid-edge cells, relative to the field RMS, increases at each refinement: 24, 44 and 82 at $\Re=10$, and 17, 25 and 41 at $\Re=997$. The low-Reynolds sequence nearly doubles; the high-Reynolds ratios are 1.47 and 1.64. These finite-grid observations are consistent with a singular edge response, but do not establish an asymptotic rate for the normalized pressure at high Reynolds number.
\item \emph{Energy shares.} The share of a band of fixed physical width grows under refinement (66.6\%, 72.5\%, 76.6\% at $\Re=10$), as the logarithmic divergence of the energy predicts. The share of a band of fixed cell count falls (83.9\%, 72.5\%, 62.7\%). Neither is resolution independent, so the 69\% of \cref{sec:where} is a $128^3$ number.
\item \emph{The residual $\delta B$.} It decreases only slowly with the unregularized lid, from 5.5\% to 3.9\% of the energy at $\Re=10$.
\item \emph{Regularized lid.} The edge pressure appears to stabilize over the tested grids (12.8, 15.0, 16.0 at $\Re=10$); the band of physical width $1/16$ holds 49--61\% of the energy. The residual energy ratio decreases (0.36\%, 0.06\%, 0.01\%). This evidence concerns the specified regularization profile and measured grids, not a proof of the continuum limit for every regularized lid.
\end{itemize}

\begin{table}[!htb]
\centering\small
\caption{\textbf{Regularized lid, cohort A} (60 cases at $128^3$; cohort-mean relative errors and recall, \%; \status{retrospective}). $s_{\mathcal J}$: truth energy share of the junction. ``+ reference'' adds $\PiF\delta B_{128}$ from a reference refitted to regularized empty-cavity solutions. $e_{\mathcal J}$: the junction's share of the arm's squared error. The surrogate is unchanged and was trained on unregularized-lid data.}
\label{tab:reglid}
\setlength{\tabcolsep}{3.5pt}
\begin{tabular}{@{}llrrrrrrrrrr@{}}
\toprule
Lid & Arm & $s_{\mathcal J}$ & $E_{\mathrm{vol}}\!\downarrow$ & $E_{\mathrm{near}}\!\downarrow$ & $E_{\nabla}\!\downarrow$ & $E_{\mathrm{pk}}\!\downarrow$ & $E_{\mathrm{obj}}\!\downarrow$ & $R_{\mathrm{obj}}\!\uparrow$ & $E_{\mathcal J}$ & $E_{\mathcal I}$ & $e_{\mathcal J}$ \\
\midrule
sharp & lifted & 68.6 & 20.6 & 10.1 & 53.5 & 21.8 & 8.4 & 86.0 & 24.4 & 6.9 & 95.1 \\
 & + reference &  & 7.5 & 9.6 & 12.5 & 7.2 & 8.4 & 86.0 & 7.8 & 6.9 & 75.4 \\
regularized & lifted & 51.5 & 31.4 & 12.7 & 126.5 & 34.6 & 9.8 & 85.5 & 43.3 & 9.1 & 93.5 \\
 & + reference &  & 32.6 & 12.8 & 131.8 & 36.0 & 9.8 & 85.5 & 45.1 & 9.2 & 93.9 \\
\bottomrule
\end{tabular}

\end{table}

With the regularized lid (\cref{tab:reglid}), the junction still holds 51.5\% of the truth energy, against 68.6\% with the unregularized lid. The lid-edge gradients remain steep even without a singularity. For the frozen sharp-lid-trained surrogate, the reference correction does not help ($\Evol$ 31.4\% $\rightarrow$ 32.6\%). Both the target boundary condition and the reference change in this diagnostic; it does not isolate the effect for a matched regularized-lid-trained model. A smaller $\delta B_{128}$ limits its possible effect but does not determine its sign. The surrogate, trained on unregularized-lid data, now errs mostly at the lid edges: 93.5\% of its squared error lies in $\mathcal J$. The truth-energy concentration motivates regional reporting under both tested lid profiles. The 93.5\% error share additionally reflects a boundary-condition distribution shift, and cannot establish a model-independent regularized-lid error mechanism. The measured benefit of $\delta B_{128}$ is specific to the tested sharp-lid composition and grid pair.

\paragraph{More tightly converged test truth.} A predeclared experiment re-solved 12 cohort-A cases, six geometries at $\Re=10$ and 900, starting from the terminal state of the test truth (\cref{app:h6}, \cref{tab:h6}). The first arm, A$^+$, keeps the discretization and tightens the linear-solver tolerance to $10^{-11}$ and the residual controls to $10^{-8}$ (pressure) and $10^{-9}$ (velocity). The second, B, additionally replaces the convection scheme by \texttt{limitedLinear}. For \HsixScope, A$^+$ converged after \HsixItMin--\HsixItMax{} additional iterations and then changed by at most $\HsixDrift$ over 100 further iterations. It differs from the test truth by at most $\HsixDiscGlobal$ globally and $\HsixDiscNear$ near the bodies. No cohort mean of any metric of ours, the lifted surrogate or the no-learning control moves by more than a relative $\HsixMetric${}\HsixGuard. B converged in none of these cases within 16{,}000 additional iterations: its pressure residual stalled at $\HsixBpTenLo$--$\HsixBpTenHi$ at $\Re=10$ and at $\HsixBpNineLo$--$\HsixBpNineHi$ at $\Re=900$. By the predeclared rule the scheme comparison is therefore unresolved. Tightening test-truth convergence leaves the reported metric effects essentially unchanged, but sensitivity to the convection scheme could not be evaluated from converged alternate-scheme fields.

\subsection{Cost}\label{sec:cost}
With 48 CPU threads the full predictor needs a median 0.76\,s per case after a warm-up call. The surrogate takes 0.45\,s, including feature and surface-graph construction, and lift, reference and centering take 0.32\,s. Thirty-six replays reproduce the sealed arrays exactly. The native reference CFD took a median 12.7\,h (cohort A) and 17.0\,h (cohort B) of single-thread wall time per case, with 7{,}205--13{,}367 SIMPLE iterations. The method's offline costs are the 1{,}620 training solves at $64^3$, the 17 empty-cavity solves at $128^3$ and training. These numbers are indicative and do not support a matched-resource speed-up claim.

{\let\FloatBarrier\relax
\section{Discussion}\label{sec:discussion}}

\paragraph{Implications for benchmarking.} \Cref{eq:decomposition} holds for any model, but distinguishes truth-energy weights from model-dependent error shares. In our cavity the junction truth-energy share ranges from 66.6\% to 69.8\% across the tested Reynolds numbers at $128^3$ (overall mean 68.6\%). For the tested transferred coarse model, the global error primarily reflects its lid-edge error; this is not a universal consequence of the truth weights. Two models can differ by a factor of 2.75 in global error and not at all in the interior. A body-free field can match a body-aware predictor globally, and an improvement near bodies may barely move the global number. The share grows with refinement for a band of fixed physical width and is still about one half with a regularized lid (\cref{sec:physics}). We therefore recommend that cavity pressure benchmarks:
\begin{enumerate}[label=(\roman*)]
\item report the truth's junction energy share $s_{\mathcal J}$ and each model's error share $w_{\mathcal J}$ at each Reynolds number and resolution, specifying the band width in both cells and physical units; when comparing grids, report a fixed physical width as well as a fixed cell count;
\item when $s_{\mathcal J}$ is large, report errors in the junction, the rest of the wall shell and the interior separately, and near-body errors outside the shell;
\item report near-body errors also relative to the body-induced field $\PiF[p-B]$, where an empty-cavity solution $B$ is available;
\item state whether and how the lid is regularized.
\end{enumerate}
The partition requires only masks and regional squared-error sums, which our analysis scripts compute.

\paragraph{What the composition is, and is not.} The reference correction reuses offline CFD to supply a body-independent, resolution-specific component that is lost when a coarse field is transferred to the native grid. It is useful when:
\begin{itemize}
\item a learned model must run on a finer grid than it was trained on;
\item the boundary singularity is shared across queries;
\item a small library of body-free solutions is affordable.
\end{itemize}
It is not a super-resolution method: the learned field is transferred, not refined, and nothing new is learned near the bodies. It is not the best use of the reference either: rebasing the surrogate on it does better on every global metric (\cref{sec:rebase}). It is also not portable by construction. Its target is a solver- and grid-specific boundary field (\cref{tab:grid}); the corresponding restriction--prolongation residual decreases over the measured regularized-lid refinements.

\needspace{5\baselineskip}\paragraph{Limitations.}
\begin{itemize}
\item \emph{One configuration.} One cavity, lid speed, solver and grid pair, tested only at the five training Reynolds numbers.
\item \emph{Numerical references.} Grid dependence was measured on the empty cavity, not on the test cohorts, whose $256^3$ solutions would each cost weeks of CPU time. At the lid edges the unregularized fields depend on the grid by nature. Tighter test-truth convergence has negligible metric effects on the 12 tested cases, but the only alternative convection scheme we tried did not converge, so scheme sensitivity remains unresolved.
\item \emph{Few independent units.} Twelve geometries per cohort, from one procedural generator and three families. Only one test geometry had a material share of its near-body band in the lid-edge junction.
\item \emph{Controls on inspected data.} The no-learning control, the matched FNO, the seed replication, the diagnostic baselines and the regularized-lid analysis were run after the truth had been inspected. The regularized-lid comparison uses a surrogate trained on unregularized-lid data. The FNO is one untuned configuration, and no native-$128^3$-trained baseline was run; the 95\% junction error share is specific to transferred coarse models.
\item \emph{Open object-local problem.} Object-local errors at high Reynolds number remain unsolved. Our volumetric near-body metrics are not surface pressure or force, and the AI-agent visual reviews were not blinded or a substitute for human validation.
\end{itemize}

\section{Conclusion}
For the tested $64^3\!\to\!128^3$ transferred surrogate and sharp-lid OpenFOAM reference fields, most of the pressure squared norm and surrogate error sits in 1.5\% of the fluid at the lid edges, predominantly at the two edges where the lid scrapes against a side wall. Adding a body-free, Reynolds-conditioned reference to the frozen surrogate lowers mean full-volume relative $L_2$ by 63.5\% (20.6\% to 7.5\%) in the prospective comparison on 12 unseen geometries. A post hoc reconstructed contrast on 12 more shows a similar global reduction. In the matched control, the learned term carries the substantial interior and near-body improvement. A body-free field alone scores about as well globally as the body-aware predictor, and rebasing on the reference lowers the global error further with little change in cohort-mean body metrics. Separating the two makes the evaluation interpretable. The junction truth-energy share depends on the lid profile and band definition: it grows under refinement at fixed physical width, decreases at fixed cell count, and remains substantial for the tested regularized profile. The reported metric effects are insensitive to tighter test-truth convergence in the 12 tested cases. The headline improvement is a lid-edge correction, and the unsolved problem is the pressure structure around bodies at high Reynolds number.

\FloatBarrier
\subsubsection*{Broader Impact Statement}
This work concerns surrogate models for laminar incompressible flow in a canonical configuration. We are not aware of direct negative societal impacts. Our main practical message is cautionary: a global error metric can overstate how well a surrogate captures the physics near objects. Surrogate predictions should not replace validated simulation in engineering decisions without region-resolved checks.

\ifdefined\DThirtyNineNamedDraft
\subsubsection*{Funding and Competing Interests}
This work was funded by EXDYNA Pte. Ltd. Several authors are affiliated with EXDYNA Pte. Ltd., as listed on the first page. The authors declare no other competing interests.
\fi

\subsubsection*{Reproducibility Statement}
The supplementary material contains:
\begin{itemize}
\item the geometry recipes of the 12 cohort-A test geometries;
\item a cohort-A score-replay package containing four-arm metric means, four-arm per-case volume errors, the two prospectively compared arms' per-case metrics, and signed regional squared-error savings for the reference and learned-term contrasts. A complementary accounting file provides all 540 case--region records, including cell counts, truth and four-arm error squared norms, body-response denominator energies and source identities; its script reconstructs regional means and separately labeled arithmetic/geometric-ratio statistics.
\item for the diagnostic baselines and the numerical-robustness study, the per-case result records, the protocols frozen before the runs, the solver and analysis scripts and a script that regenerates the corresponding tables;
\item manifests linking the packaged records and replay outputs to their source identities and file hashes, together with dated original prediction-seal and truth-attempt metadata. These are author-maintained protocol records, not independently certified timestamps; hashes alone do not establish temporal order.
\end{itemize}
The replay packages verify the included scores, spatial accounting and numerical-study tables, and rebuild the score-only effect plot (\cref{fig:effects}); they do not regenerate field figures or reproduce field inference, training or CFD solutions. Model weights, CFD arrays and training code are not included in the submission. \Cref{app:implementation,app:reference,app:cfd,app:physics,app:h6} describe the architecture, training, reference fit, geometry construction, solver settings and the new runs.

\bibliography{references}
\bibliographystyle{tmlr}

\clearpage
\appendix
\section{Metric, transfer and decision-rule details}\label{app:metric}

\paragraph{Centering and supports.} Prediction and truth are centered separately on the same mask $m=\mathbf 1[\mathrm{SDF}>0]$ and zeroed outside the fluid, so a constant shift of either raw field leaves every metric unchanged. The near-body band $\mathcal N$ is open at $\mathrm{SDF}=0$ and closed at $\mathrm{SDF}=0.2$ (unit-cube units). The object-local region further excludes three cells at each outer face; this margin is defined in cells, so its physical thickness differs between $64^3$ and $128^3$.

\paragraph{Gradient error.} For axis $a$, let $V_a$ contain cell centers whose center and both axis neighbours are fluid, excluding the two end planes of that axis. With $d_aq_i=(q_{i+e_a}-q_{i-e_a})/2$,
\begin{equation}
\Egrad=\Big[\textstyle\sum_a\sum_{i\in V_a}(d_a\hat p_i-d_ap_i)^2\big/\sum_a\sum_{i\in V_a}(d_ap_i)^2\Big]^{1/2}.
\end{equation}
This pools energy across axes. It is a relative discrepancy of discrete gradients, not an equation residual. It includes the wall shell: only one stencil cell is eroded.

\paragraph{High-magnitude regions and recall.} A region with $N\geq20$ eligible cells selects $k=N-\lfloor0.95N\rfloor$ cells by stable descending sort of $|p|$ in C-order (ties resolved by the lower flattened index). A flat prediction selects an empty set and has zero recall. A flat truth region, with range $\leq10^{-12}\max(1,\max|p|)$, is invalid. Relative-field norms $\leq10^{-15}$ and gradient squared norms $\leq10^{-24}$ are rejected. $\Robj$ measures overlap of absolute-pressure supports, not local-maximum detection or signed-lobe agreement; $\Eobj$ measures the signed field error on the truth-selected cells.

\paragraph{Transfer.} For a source cell-centered field $q$ with source fluid support $m_s$ and target support $m_t$,
\begin{equation}
L_m(q)_j = \begin{cases} I(m_s q)_j/I(m_s)_j,&m_{t,j}=1\ \text{and}\ I(m_s)_j>10^{-6},\\0,&\text{otherwise,}\end{cases}
\end{equation}
where $I$ is trilinear interpolation at source-index coordinates $(j+\tfrac12)n_s/n_t-\tfrac12$ with nearest extension at the domain boundary. A constant source value is reproduced on supported cells. The transfer is rejected when fewer than $\lfloor0.99|\mathcal F_t|\rfloor$ target cells are supported. Accumulation is float64 and output float32. $R_2$ is $2^3$ block averaging, so \cref{eq:subgrid} is a restriction--prolongation residual, not a spectral projection; $R_2\delta B=0$ is not assumed.

\paragraph{Percentiles.} p90 and p10 use linear interpolation at zero-based indices $0.9(n-1)$ and $0.1(n-1)$ of the sorted case scores (NumPy's default).

\FloatBarrier
\subsection{Predeclared decision rule for cohort A}\label{app:criteria}
Before the cohort-A truth was generated we fixed the following criteria.
\begin{itemize}
\item \emph{Absolute thresholds} on the reference-corrected predictor. Maximum mean/p90 errors: $.12/.18$ for $\Evol$, $.25/.40$ for $\Enear$, $.18/.28$ for $\Egrad$, $.12/.20$ for $\Epk$ and $.25/.40$ for $\Eobj$. Minimum recall mean/p10: $.65/.35$. All were met (\cref{tab:tails}).
\item \emph{Relative to the lifted arm}, cohort level. At least 10\% mean reduction in $\Evol$, $\Egrad$ and $\Epk$. At most 2\% mean increase in $\Enear$ and $\Eobj$. At most $.01$ absolute loss in $\Robj$.
\item \emph{Relative to the lifted arm}, per geometry, family and Reynolds group. A joint condition: lower mean $\Evol$, at most 2\% higher $\Enear$ and $\Eobj$, and at most $.01$ lower $\Robj$. Required coverage: at least 9/12 geometries, 2/3 families and 4/5 Reynolds groups.
\item \emph{Visual check}: shared-scale inspection for new artifacts such as frames, checkerboards, halos, displaced geometry or amplitude collapse.
\end{itemize}

\begin{table}[!htb]
\centering\small
\caption{\textbf{Distribution of the reference-corrected predictor's scores} (60 cases per cohort, \%; prospective absolute scores). For recall the tail is the 10th percentile and the worst is the minimum.}
\label{tab:tails}
\begin{tabular}{@{}ll*{6}{r}@{}}
\toprule
Cohort & Statistic & $E_{\mathrm{vol}}\!\downarrow$ & $E_{\mathrm{near}}\!\downarrow$ & $E_{\nabla}\!\downarrow$ & $E_{\mathrm{pk}}\!\downarrow$ & $E_{\mathrm{obj}}\!\downarrow$ & $R_{\mathrm{obj}}\!\uparrow$ \\
\midrule
A & Mean & 7.5 & 9.6 & 12.5 & 7.2 & 8.4 & 86.0 \\
A & p90 (recall: p10) & 8.4 & 12.3 & 14.9 & 8.5 & 13.5 & 67.1 \\
A & Worst & 9.2 & 20.2 & 15.1 & 8.6 & 22.8 & 41.5 \\
\midrule
B & Mean & 7.4 & 9.8 & 12.5 & 7.3 & 9.7 & 83.3 \\
B & p90 (recall: p10) & 8.3 & 13.7 & 14.9 & 8.5 & 16.2 & 65.3 \\
B & Worst & 8.5 & 15.2 & 15.2 & 8.6 & 19.7 & 43.6 \\
\bottomrule
\end{tabular}

\end{table}

The joint condition held in 11/12 geometries, 3/3 families and 5/5 Reynolds groups, and no new artifact was found. These are application-specific performance criteria, not significance tests.

\section{Coarse surrogate: specification}\label{app:implementation}
\begin{figure}[!tbp]
\centering
\includegraphics[width=399pt]{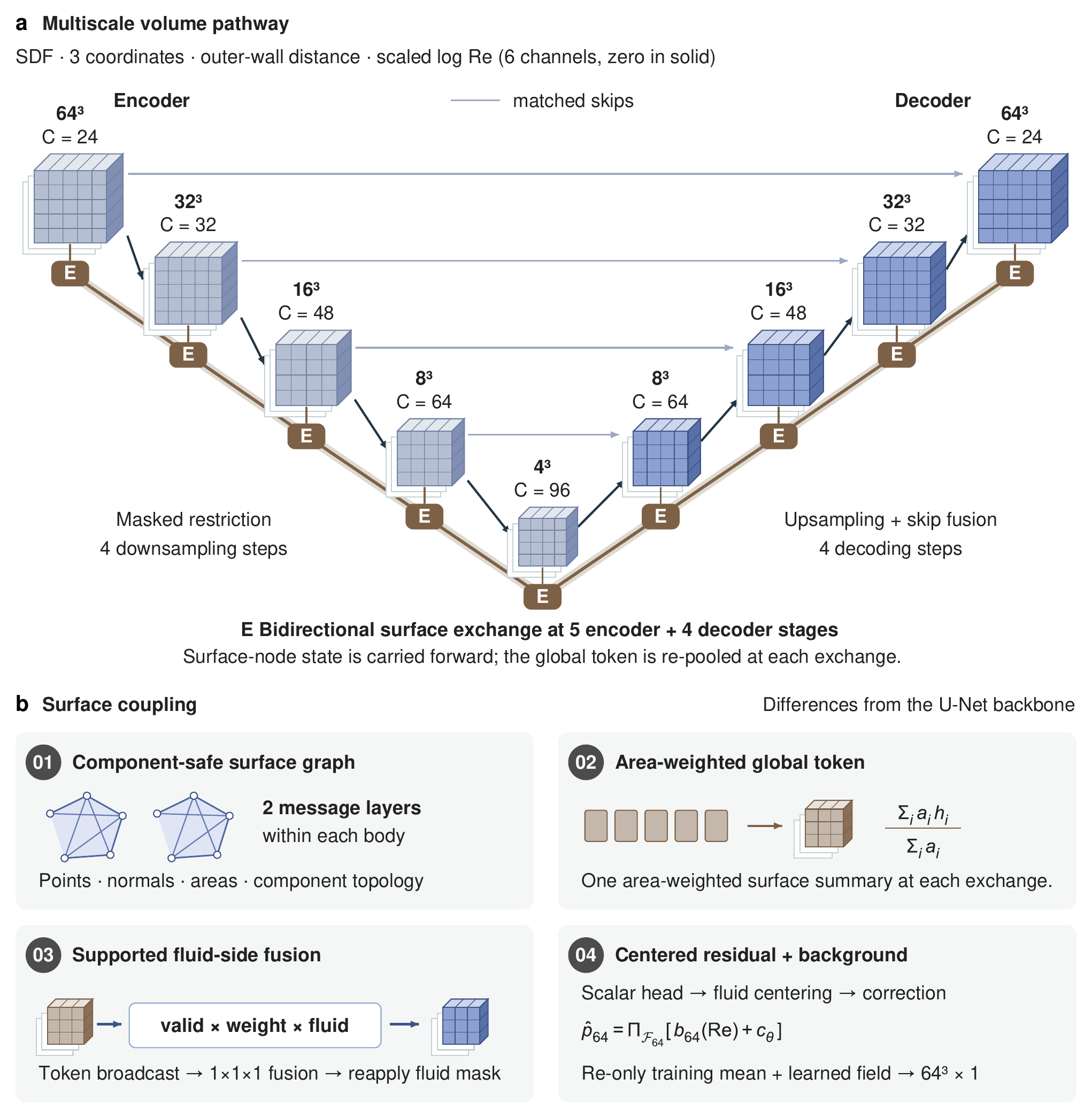}
\caption{\textbf{Coarse surrogate architecture} (grid and channel labels are exact; cube sizes are schematic). \textbf{a}, Multiscale volume pathway: a 3D U-Net-style encoder (light blue) and decoder (darker blue) joined by matched skips (thin lines) \citep{unet,unet3d}. Brown marks the surface pathway, including the nine bidirectional volume--surface exchanges (E). \textbf{b}, The four surface-coupling components that differ from the U-Net backbone; the numbers only index the boxes. The learned correction $c_\theta$ is added to the Reynolds-only background $b_{64}(\Re)$ and centered on the $64^3$ fluid by $\Pi_{\mathcal F_{64}}$.}
\label{fig:r0}
\end{figure}

\paragraph{Inputs and background.} At unit-cube cell center $x$ the six dense channels are:
\begin{itemize}
\item $\mathrm{SDF}/\sqrt3$;
\item the three components of $2x-1$;
\item twice the nearest outer-wall distance;
\item $r_*=2(\log_{10}\Re-1)/(\log_{10}900-1)-1$, which maps $\Re=10$ to $-1$ and $\Re=900$ to $+1$.
\end{itemize}
All channels are zeroed where $\mathrm{SDF}\leq0$. For family $f$ and Reynolds number $r$, with 54 training cases $T_{f,r}$, fluid masks $m_k$ and centered pressures $p_k^c$,
\begin{equation}
b_{f,r}(i)=\frac{\sum_{k\in T_{f,r}}m_k(i)p_k^c(i)}{\sum_{k\in T_{f,r}}m_k(i)},\qquad b_{64}(r)=\tfrac16\textstyle\sum_{f=1}^{6}b_{f,r},
\end{equation}
with zero at unsupported cells. No family label is used at inference.

\paragraph{Objective.} With $p^c=\Pi_{\mathcal F_{64}}p$ and $\hat p^c=\Pi_{\mathcal F_{64}}(b_{64}+c_\theta)$, the loss is
\begin{equation}
\mathcal L=\mathcal L_{\mathcal F}+\mathcal L_{\mathcal N}+\mathcal L_{\mathrm{pk}}+0.1\,\mathcal L_{\nabla},\qquad
\mathcal L_A=\frac{\sum_{i\in A}(\hat p_i^c-p_i^c)^2}{\sum_{i\in A}(p_i^c)^2}.
\end{equation}
The peak support is $|p^c|$ at or above the 95th percentile within $\mathcal N$, after excluding three outer-wall cells. $\mathcal L_\nabla$ is the pooled gradient ratio defining $\Egrad^2$. Losses compare complete centered fields, not the neural residual alone.

\paragraph{Optimization.} Training uses the following settings:
\begin{itemize}
\item 1{,}620 cases from 324 geometries, with 40{,}500 single-case updates;
\item AdamW with learning rate $3\times10^{-4}$, weight decay $10^{-5}$ and per-update cosine decay;
\item bf16 autocast, gradient clipping at 1.0 and activation checkpointing;
\item seed 42 for the sealed run;
\item the terminal checkpoint fixed in advance, with no validation-based selection and no early stopping.
\end{itemize}

\paragraph{Volume pathway.} A $1^3$ input projection, one residual block, four restriction stages (encoder widths 24, 32, 48, 64, 96), two bottleneck blocks and four upsampling stages with skip connections. Each residual block has two $3^3$ support-normalized convolutions,
\begin{equation}
\big[(K*(s v))\cdot 27/\max(\mathbf 1_{3^3}*s,1)+b\big]\,\mathbf 1[s>0],
\end{equation}
each preceded by channelwise LayerNorm, with SiLU and a residual multiplier initialized at 0.1. Restriction returns $\overline{sv}/\max(\bar s,10^{-6})$. A zero-initialized $1^3$ head gives $c_\theta$ after fluid masking and centering.

\paragraph{Surface pathway.} Marching cubes on the sampled $64^3$ SDF gives triangles, grouped by connected solid component and nearest fluid owner cell into area-weighted surface nodes with 13 features (position, normal, log-area, wall distance, offset from the component centroid, component area fraction, $r_*$). Graph neighbours are the 12 nearest nodes of the same component. At each of nine exchanges (five encoder, four decoder), two message-passing layers update a width-48 surface state. The layers use the edge features $[h_i, h_j-h_i, \delta_{ij}, \|\delta_{ij}\|, n_i\cdot n_j, \delta_{ij}\cdot n_i, \delta_{ij}\cdot n_j]$. The state then gathers volume features at owner cells. In the reverse direction, node states are area-averaged into one global token. The token is fused into active cells within unit-cube distance 0.25 of the surface, with weight $(1-d/0.25)^2$. \Cref{fig:r0} shows the architecture. The network has 3{,}150{,}088 parameters.

\section{Reference fit and fidelity}\label{app:reference}
Each of the 17 empty-cavity pressures, at $\Re=$ 10, 13, 18, 24, 32, 42, 56, 75, 100, 133, 177, 237, 316, 421, 561, 748 and 997, is centered, and the ensemble mean $\mu$ is subtracted. For $X=U\Sigma V^\top$ the first eight rows of $V^\top$ give orthonormal modes $\phi_j$ and $U_{kj}\Sigma_{jj}$ the training coefficients. Reconstruction is
\begin{equation}
B_{128}(\Re)=\Pi_{\mathrm{cavity}}\Big[\mu+\textstyle\sum_{j=1}^8 a_j(\log_{10}\Re)\,\phi_j\Big],
\end{equation}
with shape-preserving piecewise-cubic interpolants $a_j$ and no extrapolation. Mean and basis are stored in float32 and coefficients are accumulated in float64. Rank eight was fixed before any test outcome. The fit was validated \emph{before} this study on 16 held-out native-grid empty-cavity solutions at $\Re=11,15,\dots,864$ (\cref{tab:reference}). Visual inspection of the three worst cases ($\Re=864, 365, 648$) found no artifacts. Of the five test Reynolds numbers, 30, 300 and 900 are interpolated and 10 and 100 are fit points. At the 17 fit points the rank-8 reconstruction differs from the library fields by at most $3.5\times10^{-4}$ in relative $L_2$ norm ($\Re=561$), so fit points are reproduced closely but not exactly.

\section{Geometry construction, CFD settings and cost}\label{app:cfd}

\paragraph{Geometry.} Frames are sampled from normalized Gaussian directions with Gram--Schmidt orthogonalization.
\begin{itemize}
\item \emph{Capsules} join a cylinder of length $U(0.18,0.28)$ and radius $U(0.055,0.075)$ to same-radius end spheres.
\item \emph{Pairs} combine a sphere of radius $U(0.060,0.085)$ with an oriented box whose half-sizes lie between $(.045,.050,.055)$ and $(.070,.080,.090)$. The center separation is the sphere radius plus the box bounding radius plus $U(.025,.055)$.
\item \emph{Triplets} place a sphere ($U(.050,.065)$), a cylinder (radius $U(.035,.045)$, length $U(.10,.14)$) and a box (half-sizes from $(.035,.040,.045)$ to $(.050,.055,.060)$) at $-s,0,+s$ along a frame axis.
\end{itemize}
Placement strata constrain the bounding-sphere gap to the selected face to $[0.025,\min(0.07,0.48-r_b)]$. Generator seeds are 2026090901 (cohort A) and 2026091407 (cohort B). All 24 geometries are disjoint by hash from all training and development geometries. Supplementary Data S1 lists the 12 accepted cohort-A recipes. SDFs are unions of exact primitive SDFs without smoothing. The box SDF is $\phi(|R^\top(x-c)|-h)$ with $\phi(v)=\|v_+\|_2+\min(\max_jv_j,0)$, and cylinders have flat caps.

\paragraph{Solver settings.} All 120 test cases were rendered from one solver-control template (SHA-256 prefix \texttt{4177e7a6}). The settings are:
\begin{itemize}
\item \texttt{steadyState}; \texttt{Gauss linear} gradients; \texttt{bounded Gauss linearUpwind grad(U)} convection; \texttt{Gauss linear corrected} Laplacians;
\item GAMG (pressure) and \texttt{smoothSolver} (velocity) with Gauss--Seidel smoothing, tolerance $10^{-9}$ and relative tolerance 0.01;
\item SIMPLE with no non-orthogonal correctors, relaxation 0.3/0.5 and residual controls $10^{-6}/10^{-7}$;
\item zero initial fields and zero-gradient pressure on all walls; the lid velocity is not smoothed at the edges.
\end{itemize}
Acceptance required reported convergence within 16{,}000 iterations, normal termination and finite continuity diagnostics. After stripping comments, \texttt{fvSchemes} and \texttt{fvSolution} are byte-identical between the test cases and all 33 native empty-cavity cases, i.e.\ the 17 library cases and the 16 held-out cases of \cref{tab:reference}. The library's \texttt{controlDict} differs only in restarting from the latest time after an iteration-ceiling extension and in writing six rather than twelve significant digits. We re-verified the $64^3$ cases retrospectively: of all 2{,}874 $64^3$ cases on the compute host, 2{,}863 have \texttt{fvSchemes} and \texttt{fvSolution} identical to the test template after comment stripping. The other 11 are solver diagnostics from two development experiments, not training cases; they differ only in residual tolerances tightened to $10^{-9}$ and $10^{-10}$. Terminal global continuity errors of the cohort-A truth are at most $2.7\times10^{-18}$ in magnitude. Failed-job and retry histories are incomplete; every prespecified test case is reported.

\paragraph{Cost.} Single-thread \texttt{simpleFoam} wall time per test case was 6.9--16.3\,h (median 12.7\,h) in cohort A and 9.5--20.3\,h (median 17.0\,h) in cohort B, with 48 and 60 concurrent single-thread jobs, respectively. Cases took 7{,}205--13{,}367 iterations. Surrogate timing is given in \cref{sec:cost}.

\section{Least-squares representation residual and diagnostic baselines}\label{app:floor}
\paragraph{Least-squares representation residual.} For a centered field $q$ on the $128^3$ fluid mask, the target optimization is
\[
\eta_*(q)=\min_{u\in\mathbb R^{|\mathcal F_{64}|},\,c\in\mathbb R}\ \|q-L_m u-c\,\mathbf 1\|_{2,\mathcal F}\,\big/\,\|q\|_{2,\mathcal F}.
\]
We report residuals attained by LSQR using the sparse matrix of the mask-normalized transfer $L_m$ (\cref{app:metric}), which reproduces the transfer of the paper to $1.2\times10^{-7}$. LSQR uses $\mathtt{atol}=\mathtt{btol}=10^{-10}$ and at most 4{,}000 iterations. Convergence diagnostics were not retained, so these estimates are not certified optima. The attained residuals upper-bound $\eta_*(q)$; they do not certify a lower bound on achievable error and apply only to this transfer, not to all $64^3$ representations or decoders.

For the cohort-A truth the numerical residual is 13.8\% on average (cohort B: 13.8\%). Evaluating the same globally fitted residual within each region gives case-mean relative norms of 16.6\% in $\mathcal J$, 14.8\% in the wall shell, 3.3\% on the object-peak support, 2.9\% in the interior near-body band and 1.0\% in the interior. These are not separately optimized regional lower bounds. The case-mean share of residual energy in $\mathcal J$ is 99.9\%; this is not the fraction of truth energy lost. For the empty-cavity reference ($\Re=10,30,100,300,900$), the attained residuals are 17.1\%, 17.0\%, 15.7\%, 11.4\% and 8.4\%, against 21.2\%, 21.1\%, 19.5\%, 14.2\% and 10.6\% for $\|\delta B_{128}\|/\|B_{128}\|$. These smaller feasible residuals show that block averaging followed by the lift is not optimal; its residual norms are about a quarter larger than the attained LSQR residuals, without certifying the unrepresentable part.

\paragraph{Voxelization.} Between the $64^3$ and $128^3$ masks of the same geometry, the combined count of fine-fluid cells with a solid parent and fine-solid cells with a fluid parent is at most 0.064\% of the fine-fluid count. On cohort B the fine mask regenerated from the geometry recipe equals the truth mask in every case.

\paragraph{Diagnostic baselines.} The rebased surrogate of \cref{eq:rebase} uses $64^3$ empty-cavity solutions at the five test Reynolds numbers. They were computed with the reference library's case files, the mesh changed to $64^3$, and all converged within 3{,}643 iterations. In cohort B the lifted arm is the reconstruction $\PiF[\hat p_{128}-\PiF\delta B_{128}]$. Applied to five cohort-A cases, where the lifted arm was sealed, this reconstruction reproduces the sealed arrays to a relative error of at most $4.2\times10^{-8}$. For $e=\hat p-p$ and the set $\mathcal E_{\mathcal F}$ of interior faces joining two fluid cells, the face error is the relative norm
\[
E_{\mathrm{face}}=\left[\frac{\sum_{(i,j)\in\mathcal E_{\mathcal F}}(e_i-e_j)^2}{\sum_{(i,j)\in\mathcal E_{\mathcal F}}(p_i-p_j)^2}\right]^{1/2}.
\]
The restriction--transfer residual energy ratio is $\|e-I R_2e\|_2^2/\|e\|_2^2$, evaluated on the full grid with solids zero-filled, where $R_2$ is $2\times2\times2$ block averaging and $I$ is ordinary trilinear interpolation with nearest boundary extension, without mask normalization. Since $I R_2$ is not an orthogonal projection, this ratio is neither a frequency-band energy fraction nor a checkerboard-only fraction.

\section{Grid and regularized-lid study: protocol and runs}\label{app:physics}
The protocol was frozen and committed before any run. Every case copies an existing, validated case (mesh, schemes, linear solvers and residual controls) and changes only the lid profile or the mesh resolution; the regularized profile is evaluated at the lid face centres. Iteration ceilings were 30{,}000 ($64^3$, $128^3$) and 40{,}000 ($256^3$); convergence status is reported and no run was dropped or repeated after its result was seen.
\begin{itemize}
\item \emph{Regularized cohort A}: 60 cases, 7{,}511--13{,}645 SIMPLE iterations, all converged.
\item \emph{Regularized reference library}: 17 empty-cavity cases at the library Reynolds numbers, 7{,}640--12{,}713 iterations, all converged; refitted with rank 8 and the unchanged interpolation.
\item \emph{Grid study}: empty cavity at $\Re=10,100,997$, both lids. $64^3$: 2{,}580--3{,}592 iterations. $256^3$: eight MPI ranks, initialized by consistent field mapping from the converged $128^3$ solution of the same lid and Reynolds number, then 4{,}285--11{,}522 iterations to the unchanged residual criteria. Solving $256^3$ from rest was measured at about 22\,s per iteration, which is why the initialization was changed before any $256^3$ result existed.
\end{itemize}
The regional analysis of the regularized cohort uses the unchanged predictor, lift, centering, metric functions and regions. A self-check that applies the same code to the unregularized cohort reproduces the frozen reference fit to $1.9\times10^{-6}$ and the sealed predictions to $3.8\times10^{-6}$.

\section{Numerical-reference study: protocol and runs}\label{app:h6}
Cases, arms, thresholds and the decision precedence were frozen before any run. The 12 cases are capsules 00 and 01, pairs 00 and 02 and triplets 00 and 03 (in each family the interior placement and one other placement stratum), each at $\Re=10$ and 900. Both arms start from independent copies of the terminal $U$, $p$ and $\phi$ of the test truth A on the same native mesh and may use at most 16{,}000 additional SIMPLE iterations.
\begin{itemize}
\item \emph{A$^+$}: the discretization of A; linear-solver tolerance $10^{-11}$ with zero relative tolerance; residual controls $10^{-8}$ (pressure) and $10^{-9}$ (velocity).
\item \emph{B}: as A$^+$, with \texttt{bounded Gauss limitedLinear 1} convection instead of \texttt{linearUpwind}.
\end{itemize}
After its strict endpoint each arm runs exactly 100 further iterations without residual control. The relative change of the centered pressure over these iterations is the drift, with validity limits $10^{-4}$ globally and $2\times10^{-4}$ near the bodies. An unconverged arm makes the comparison unresolved, and no later rule can override this. The runs used CPU jobs on Google Cloud with the pinned OpenFOAM v2412 runtime.

In the original jobs the drift endpoint was not written. After a restart OpenFOAM's time index is not aligned with the time value, and with a write interval of 100 the step strict$+100$ was skipped. The strict endpoints, meshes and dictionaries had been stored, so we repeated the drift invocation from the stored strict endpoints with a write interval of 1 and otherwise identical dictionaries. For one case we compared the repeated solver log with the original: all 100 iterations have identical residuals, and only the cumulative continuity diagnostic, which is not restored on restart, differs.

\begin{table}[!htb]
\centering\small
\caption{\textbf{Numerical-reference study} (\HsixScope; descriptive). ``A$^+$ iterations'': additional SIMPLE iterations to strict convergence. ``A$^+$ vs.\ A'': relative $L_2$ difference of the centered pressure on the full fluid and on the near-body band. ``Drift'': the same measure over 100 further iterations. Last column: initial pressure residual of B at its iteration cap, where the strict control is $10^{-8}$.}
\label{tab:h6}
\begin{tabular}{@{}lrrrrrrr@{}}
\toprule
 & & A$^+$ & \multicolumn{2}{c}{A$^+$ vs.\ A ($10^{-5}$)} & \multicolumn{2}{c}{A$^+$ drift ($10^{-8}$)} & B: $p$ residual \\
\cmidrule(lr){4-5}\cmidrule(lr){6-7}
Geometry & $\Re$ & iterations & global & near & global & near & at 16{,}000 \\
\midrule
capsule-00 & 10 & 7{,}096 & 0.32 & 0.65 & 0.18 & 0.36 & $1.9\times10^{-7}$ \\
capsule-00 & 900 & 3{,}879 & 1.17 & 1.63 & 1.75 & 3.50 & $3.3\times10^{-3}$ \\
capsule-01 & 10 & 7{,}215 & 0.56 & 1.45 & 0.28 & 0.67 & $5.5\times10^{-7}$ \\
capsule-01 & 900 & 3{,}856 & 1.67 & 2.29 & 1.70 & 2.93 & $2.9\times10^{-3}$ \\
pair-00 & 10 & 7{,}117 & 0.25 & 1.07 & 0.14 & 0.62 & $7.0\times10^{-7}$ \\
pair-00 & 900 & 3{,}785 & 1.45 & 2.21 & 1.87 & 3.37 & $2.0\times10^{-3}$ \\
pair-02 & 10 & 7{,}147 & 0.23 & 0.41 & 0.13 & 0.20 & $3.1\times10^{-7}$ \\
pair-02 & 900 & 3{,}682 & 1.12 & 1.65 & 2.51 & 3.04 & $2.6\times10^{-3}$ \\
triplet-00 & 10 & 7{,}245 & 1.01 & 3.84 & 0.49 & 1.81 & $4.5\times10^{-7}$ \\
triplet-00 & 900 & 3{,}664 & 1.44 & 1.17 & 1.83 & 1.87 & $3.2\times10^{-3}$ \\
triplet-03 & 10 & 6{,}898 & 0.50 & 2.18 & 0.24 & 1.07 & $1.9\times10^{-7}$ \\
triplet-03 & 900 & 3{,}460 & 1.44 & 2.18 & 1.88 & 3.63 & $2.5\times10^{-3}$ \\
\bottomrule
\end{tabular}

\end{table}

\section{Development history and coarse-grid evidence}\label{app:lineage}
The composition was chosen on development data, before cohort A existed. \Cref{tab:lineage} lists the distinct cohorts. None is pooled with the native-grid cohorts.

\begin{table}[!htb]
\centering\small
\caption{\textbf{Evaluation cohorts.} Cases are CFD solutions; geometries are independent units. ``Prospective'' means predictions were fixed before the reference pressure was generated. $^\dagger$For our predictor's absolute scores; the lifted contrast is \status{post hoc}.}
\label{tab:lineage}
\begin{tabularx}{\linewidth}{@{}>{\raggedright\arraybackslash}p{36mm}>{\raggedright\arraybackslash}p{22mm}rl>{\raggedright\arraybackslash}X@{}}
\toprule
Cohort & Role & Cases/geom. & Grid & Comparison \\
\midrule
Coarse replication 1 & prospective & 120/24 & $64^3$ & volume baseline (text) \\
Coarse replication 2 & prospective & 120/24 & $64^3$ & volume and spectral baselines (\cref{tab:d17}) \\
Surface cohort & prospective & 120/24 & $64^3$ & surface-conditioned vs.\ volume model \\
Surface ablation & development & 60/12 & $64^3$ & complete vs.\ null surface input \\
Composition cohort & prospective & 80/16 & $64^3$ & coarse surrogate on new compositions \\
Freeform cohort & prospective & 80/16 & $64^3$ & coarse surrogate on freeform shapes \\
Cohort A & prospective & 60/12 & $128^3$ & this paper, main contrast \\
Cohort B & prospective$^\dagger$ & 60/12 & $128^3$ & this paper, replication \\
\bottomrule
\end{tabularx}
\end{table}

\paragraph{Coarse-grid baselines.} Four models share one coarse replication cohort (\cref{tab:d17}). All use 900 training cases, 22{,}500 updates and seed 42. The background-residual masked volume model (BR-MGL) has 2.0M parameters; the FNO has width 16, four blocks and $6^3$ modes (1.77M real parameters). Residual models predict a correction to a family/Reynolds background; direct models predict the full pressure. MGL denotes the masked volume model and BR-MGL its background-residual form. On coarse replication 1, BR-MGL gave $\Evol=5.28\%$, $\Enear=13.59\%$, $\Egrad=4.07\%$, $\Epk=2.65\%$, $\Eobj=10.45\%$ and $\Robj=80.28\%$. On separate prospective cohorts (\cref{tab:r64}), the surface-conditioned model improves on the volume model. A matched null-surface ablation on development data gives $\Enear$ 10.7\% vs.\ 12.1\%, $\Eobj$ 7.7\% vs.\ 8.8\% and $\Robj$ 83.3\% vs.\ 80.0\%. The coarse surrogate generalizes to new analytic compositions and to occupancy-derived freeform shapes at $64^3$. Its $64^3$ global error there (3.1--3.7\%), measured on its own training grid, is much smaller than the $128^3$ global error of the reference-corrected predictor (7.4--7.5\%; lifted surrogate 20.6\%), while its near-body error is similar (9.7--9.9\% vs.\ 9.6--9.8\%).

\begin{table}[!htb]
\centering\small
\caption{\textbf{Coarse-grid volume and spectral baselines} (coarse replication 2, 120 cases/24 geometries; relative errors and recall in \%). MGL: masked volume model; ``BR'' and ``residual'': correction to a family/Reynolds background; ``direct'': full pressure.}
\label{tab:d17}
\begin{tabular}{@{}lrrrrrr@{}}
\toprule
Model & $\Evol\!\downarrow$ & $\Enear\!\downarrow$ & $\Egrad\!\downarrow$ & $\Epk\!\downarrow$ & $\Eobj\!\downarrow$ & $\Robj\!\uparrow$ \\
\midrule
BR-MGL (residual) & 5.26 & 13.02 & 3.67 & 2.13 & 9.18 & 80.17 \\
Direct MGL & 10.17 & 13.03 & 15.48 & 8.38 & 9.08 & 80.13 \\
Residual FNO & 6.13 & 14.88 & 4.84 & 2.35 & 11.30 & 78.34 \\
Direct FNO & 13.04 & 16.37 & 20.97 & 10.32 & 11.38 & 77.19 \\
\bottomrule
\end{tabular}
\end{table}

\begin{table}[!htb]
\centering\small
\caption{\textbf{Separate coarse-grid prospective cohorts} (relative errors and recall in \%). Horizontal rules separate datasets. In the surface cohort, the surface-conditioned model is an earlier variant with a family/Reynolds background; in the other two cohorts, ``coarse surrogate'' is the frozen surrogate of this paper and ``background only'' its Reynolds-only background $b_{64}$.}
\label{tab:r64}
\begin{tabular}{@{}llrrrrrr@{}}
\toprule
Cohort & Predictor & $\Evol\!\downarrow$ & $\Enear\!\downarrow$ & $\Egrad\!\downarrow$ & $\Epk\!\downarrow$ & $\Eobj\!\downarrow$ & $\Robj\!\uparrow$ \\
\midrule
Surface & BR-MGL & 5.18 & 13.00 & 3.79 & 2.43 & 9.31 & 80.29 \\
 & Surface-conditioned & 4.12 & 9.81 & 3.38 & 1.83 & 6.68 & 83.95 \\
\midrule
Composition & Background only & 7.11 & 39.06 & 4.84 & 2.80 & 41.06 & 54.23 \\
 & Coarse surrogate & 3.11 & 9.68 & 2.63 & 1.24 & 7.35 & 85.94 \\
\midrule
Freeform & Background only & 9.01 & 47.59 & 5.52 & 3.78 & 48.74 & 49.62 \\
 & Coarse surrogate & 3.67 & 9.88 & 2.91 & 1.52 & 7.53 & 84.54 \\
\bottomrule
\end{tabular}
\end{table}

\paragraph{Lifting versus direct execution.} On 12 development cases (six geometries, $\Re=10,900$), the lifted surrogate had $\Evol/\Enear/\Egrad/\Eobj=19.9/11.4/49.9/9.9\%$. Executing the same weights directly at $128^3$ gave $20.6/30.7/50.0/25.7\%$, and adding the reference to the lift gave $8.4/11.4/12.0/9.9\%$. Direct execution changes the physical scale of all learned operators and produced wider body shells and displaced lobes.

\paragraph{Development energy diagnostic.} On the same development cases, the coarse-to-fine CFD discrepancy $\PiF[p^c_{128}-L_m p^c_{64}]$ had a case-mean 99.3\% of its energy in the eight-cell wall shell outside the near-body band. The reference residual $\delta B_{128}$ had more than 99.9\%. This motivated the composition; the post hoc regional analysis of \cref{sec:where} replaces it.

\paragraph{Headroom and learnability of the remaining near-body error.} Two target-informed oracle diagnostics illustrate possible near-body corrections on development and training data. Both use truth and are not predictors or optimality bounds.
\begin{itemize}
\item \emph{Learnability.} On the same 12 development cases, an oracle restricted the centered truth-minus-prediction residual of our composition to $64^3$, recentered it, lifted it back with the transfer of \cref{app:metric}, added it and recentered again. This reduced the remaining error. Residual learners using only deployment-visible inputs, with fresh or transferred initialization, did not reliably reproduce this gain. Under an equal 3{,}000-update budget, training on 24 geometries instead of a repeated set of 12 improved $\Eobj$ but still worsened $\Enear$ by 6.08\% and lowered $\Robj$ by 0.037 relative to the fixed composition.
\item \emph{Resolution.} On 120 training cases from 24 geometries at all five Reynolds numbers, both oracles add a tapered, centered truth residual $r=\PiF(p-p_b)$ to the fixed composition $p_b$: $\PiF[p_b+w\,T(wr)]$. The weight $w$ is 1 for $0<\mathrm{SDF}\le0.2$, 0 outside $(0,0.35)$, and $1-3u^2+2u^3$ in between, with $u=(\mathrm{SDF}-0.2)/0.15$. For the coarse oracle, $T$ is support-normalized restriction to $64^3$ followed by lifting. For the native oracle, $T$ is fluid-mask-normalized Gaussian smoothing at $128^3$ ($\sigma=1$ cell). The coarse oracle reaches $\Enear=2.65\%$, $\Eobj=3.19\%$ and $\Robj=98.8\%$; the native oracle reaches 2.49\%, 2.97\% and 98.8\%.
\end{itemize}
On the same data, the native-resolution oracle improves on the $64^3$ oracle by 5.9\% ($\Enear$) and 7.1\% ($\Eobj$). These two constructions attain similar near-body errors, but do not measure the fraction of all representable corrections retained at $64^3$. The first diagnostic shows that the tested learners did not reliably reproduce the target-informed gain. Representability does not establish learnability, and these unequal-degree-of-freedom constructions do not bound all representations.

\section{Additional results}\label{app:more}

\subsection{Geometry-level statistics and families}
\Cref{tab:stats} gives the post hoc geometry-level statistics behind \cref{tab:main}: medians, leave-one-geometry-out ranges, one-sided bounds and sign-flip tests. \Cref{tab:family} gives the prospective absolute scores of our predictor by family in both cohorts; object-local errors are lowest for capsules in both.

The sign-flip calculation enumerates all $2^{12}$ sign patterns of the paired geometry-mean differences. Inferential exactness requires joint sign symmetry under the null (for example, independent symmetric paired differences), not geometry independence alone. The $t$ bounds assume approximately normal geometry-level log ratios for errors or paired differences for recall. These are unadjusted, metric-wise descriptive summaries for the sampled generator and placement strata, not simultaneous guarantees or tests of the predeclared arithmetic cohort-mean thresholds.

\begin{table}[!htb]
\centering\small
\caption{\textbf{Geometry-level statistics for adding the reference} ($n=12$ geometries per cohort; cohort-B contrast post hoc). LOGO: range of the arithmetic-mean change over leave-one-geometry-out subsets. The one-sided bound is the upper 95\% bound on the geometric-mean ratio of geometry-mean errors, expressed as a relative increase ($t$ on log ratios); $^\dagger$lower 95\% bound on the mean paired recall change. Non-inferiority margins are $+2\%$ for $\Enear,\Eobj$ and $-1$\,pp for $\Robj$. In cohort A, the $\Enear$ bound exceeds its margin because of the favourable capsule-02 outlier (\cref{sec:boundary}); excluding it post hoc, the bound is below $+0.1\%$, which does not establish non-inferiority for the full cohort.}
\label{tab:stats}
\begin{tabular}{@{}llrrrrrr@{}}
\toprule
 & & Change of & Median & LOGO & One-sided 95\% & Geometries & Sign-flip \\
Cohort & Metric & means (\%) & geometry (\%) & range (\%) & bound (\%) & improved & $p$ \\
\midrule
A & $E_{\mathrm{vol}}$ & $-$63.46 & $-$63.16 & [$-$63.72, $-$63.28] & $-$62.74 & 12/12 & 0.0005 \\
A & $E_{\mathrm{near}}$ & $-$5.53 & $-$0.08 & [$-$6.11, $-$0.05] & $+$4.28 & 7/12 & 0.2168 \\
A & $E_{\nabla}$ & $-$76.59 & $-$76.70 & [$-$76.69, $-$76.55] & $-$76.39 & 12/12 & 0.0005 \\
A & $E_{\mathrm{pk}}$ & $-$66.89 & $-$66.93 & [$-$66.96, $-$66.82] & $-$66.66 & 12/12 & 0.0005 \\
A & $E_{\mathrm{obj}}$ & $+$0.08 & $-$0.09 & [$-$0.12, $+$0.13] & $+$0.95 & 8/12 & 0.9854 \\
A & $R_{\mathrm{obj}}$ (pp) & $+$0.006 & -- & -- & $-$0.014$^\dagger$ & 8/12 & 0.6108 \\
\midrule
B & $E_{\mathrm{vol}}$ & $-$64.16 & $-$64.49 & [$-$64.33, $-$64.07] & $-$63.71 & 12/12 & 0.0005 \\
B & $E_{\mathrm{near}}$ & $-$0.13 & $-$0.17 & [$-$0.17, $-$0.10] & $-$0.04 & 10/12 & 0.0513 \\
B & $E_{\nabla}$ & $-$76.69 & $-$76.73 & [$-$76.74, $-$76.67] & $-$76.57 & 12/12 & 0.0005 \\
B & $E_{\mathrm{pk}}$ & $-$66.81 & $-$66.84 & [$-$66.86, $-$66.76] & $-$66.63 & 12/12 & 0.0005 \\
B & $E_{\mathrm{obj}}$ & $-$0.10 & $-$0.07 & [$-$0.12, $-$0.08] & $-$0.04 & 10/12 & 0.0049 \\
B & $R_{\mathrm{obj}}$ (pp) & $+$0.010 & -- & -- & $-$0.007$^\dagger$ & 7/12 & 0.3232 \\
\bottomrule
\end{tabular}

\end{table}

\begin{table}[!htb]
\centering\small
\caption{\textbf{Families} (our predictor, prospective absolute scores; cohort means in \%).}
\label{tab:family}
\begin{tabular}{@{}ll*{6}{r}@{}}
\toprule
Cohort & Family & $E_{\mathrm{vol}}\!\downarrow$ & $E_{\mathrm{near}}\!\downarrow$ & $E_{\nabla}\!\downarrow$ & $E_{\mathrm{pk}}\!\downarrow$ & $E_{\mathrm{obj}}\!\downarrow$ & $R_{\mathrm{obj}}\!\uparrow$ \\
\midrule
A & capsule & 7.6 & 8.1 & 12.5 & 7.2 & 4.8 & 87.7 \\
A & pair & 7.6 & 10.7 & 12.7 & 7.2 & 10.2 & 87.8 \\
A & triplet & 7.5 & 9.9 & 12.5 & 7.3 & 10.3 & 82.4 \\
\midrule
B & capsule & 7.4 & 8.5 & 12.4 & 7.3 & 7.4 & 81.1 \\
B & pair & 7.5 & 10.8 & 12.5 & 7.3 & 11.8 & 83.7 \\
B & triplet & 7.3 & 10.0 & 12.5 & 7.2 & 10.1 & 85.0 \\
\bottomrule
\end{tabular}

\end{table}

\subsection{All case scores and linked diagnostics}
\begin{figure}[!htbp]
\centering
\includegraphics{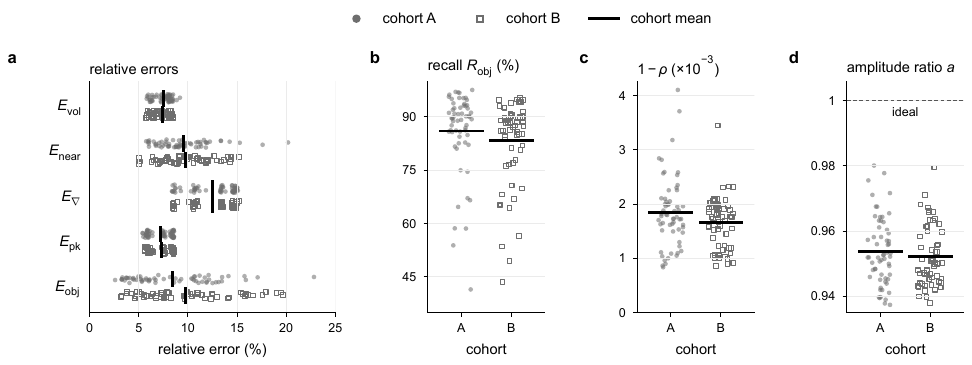}
\caption{\textbf{All 60 case scores per cohort} for our predictor (prospective absolute scores). Filled circles, cohort A; open squares, cohort B; black bars, cohort means. Points are jittered across their lane for display only. \textbf{a}, Relative errors. \textbf{b}, Peak-region recall. \textbf{c}, One minus the field correlation $\rho$. \textbf{d}, Amplitude ratio $a$, predicted over true field standard deviation; the dashed line marks $a=1$.}
\label{fig:distributions}
\end{figure}

\Cref{fig:distributions} shows every case score of our predictor in both cohorts. It includes two shape diagnostics: the correlation $\rho$ between predicted and true centered fields, and the amplitude ratio $a$, i.e.\ predicted over true standard deviation. Their cohort-A means are $\rho=0.998$ and $a=0.954$. For each case $\Evol^2=1+a^2-2a\rho$, so these are linked to $\Evol$, not independent evidence. High correlation and nearly correct amplitude coexist with the object-local errors of \cref{sec:boundary}.

\subsection{Best cases}
\begin{figure}[!htbp]
\centering
\includegraphics{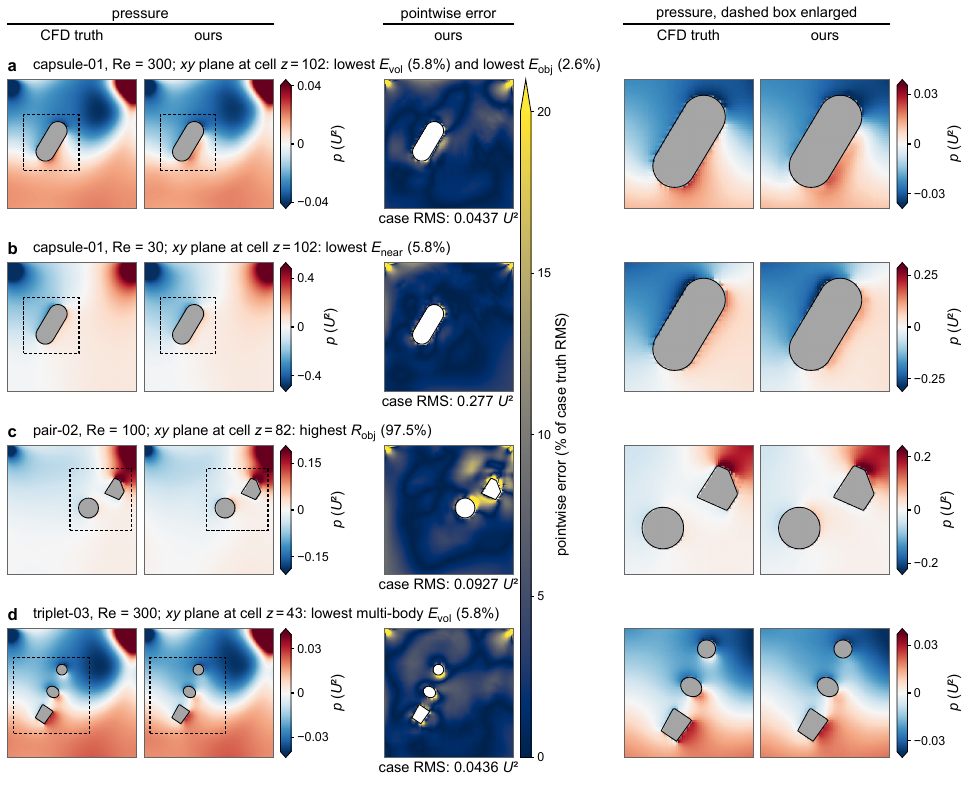}
\caption{\textbf{Best cohort-A cases} under the rule stated above: lowest $\Evol$ (also lowest $\Eobj$), lowest $\Enear$, highest $\Robj$, and lowest multi-body $\Evol$; each row shows the $xy$ plane with the most solid cells. Columns 1--2 share a symmetric scale set to the 99th percentile of $|p|$ outside the wall shell, as in \cref{fig:fields}. Column 3 is the pointwise error of our predictor as a percentage of the case's full-fluid truth RMS, with the case RMS printed below to convert to $U^2$, on the same 0--20\% scale as \cref{fig:fields}. Columns 4--5 enlarge the dashed box (all bodies plus an 8-cell margin) on the box's own symmetric scale, set to the 99.5th percentile of $|p|$ in the box. Grey is solid in pressure panels and white in error panels; black contours mark the zero level of the signed distance. Short dark marks on body outlines are first-layer fluid cells of the voxelized boundary, where the CFD pressure itself has local extremes.}
\label{fig:favorable}
\end{figure}

For balance with the failure cases of \cref{fig:fields}, \cref{fig:favorable} shows the best cohort-A cases under a fixed rule:
\begin{itemize}
\item the lowest $\Evol$, which is also the lowest $\Eobj$ (capsule-01, $\Re=300$: 5.8\% and 2.6\%);
\item the lowest $\Enear$ (capsule-01, $\Re=30$: 5.8\%);
\item the highest $\Robj$ (pair-02, $\Re=100$: 97.5\%);
\item the lowest $\Evol$ among multi-body cases (triplet-03, $\Re=300$: 5.8\%).
\end{itemize}
Shown is the $xy$ plane with the most solid cells. The selection depends on the scores, so these cases illustrate what the method does well; they do not represent the cohort.

\subsection{Gradient and peak metrics by region}
\Cref{tab:h10} repeats the regional analysis of \cref{sec:where} for $\Egrad$ and $\Epk$. It was computed by a separately frozen regional audit of the same arrays; the audit's additive identity was checked at the documented float32 tolerance, with maximum deviation $7.3\times10^{-6}$. The audit also decomposes squared errors into regional offset and within-region shape. Interior offsets account for about 1\% of interior squared error in both arms, so the interior result is not an artifact of global centering. That audit's frozen descriptive rule labels the reference gain ``spatially mixed'' rather than ``boundary dominated''. The only failing input is the pooled near-body squared-error change by Reynolds number ($-22\%$ to $-61\%$), which capsule-02's junction cells dominate. Every other input is below $10^{-4}$ against a limit of $0.05$, and excluding capsule-02 the near-body changes are within $\pm0.5\%$. We report the frozen label unchanged.

\subsection{Learned-increment accounting}
Relative to the no-learning control, the squared error that our predictor saves is split over the 12 wall-by-body regions. 64.8\% of it lies in object-local fluid and 76.2\% outside the wall shell when each case is normalized by its truth energy; pooled over cases these shares are 61.2\% and 69.5\%. Arrays reconstructed for this audit match the sealed ones to $7.3\times10^{-6}$ (pressure units), and replayed metrics match to $8.3\times10^{-16}$.

\subsection{Seed replication}\label{app:seeds}
After the method and the cohort-A result were frozen, the complete surrogate training was repeated for the predeclared seeds 42, 314159 and 271828 in a cloud environment; their evaluations on the consumed cohort A are retrospective. Architecture, data, ordering rule, objective, optimizer, budget, composition and evaluator were unchanged; only initialization, RNG and ordering seed varied. Every seed passes the predeclared absolute and learned-increment checks. The retrained seed-42 run differs from the historical seed-42 run by 0.10 points in $\Evol$ and 0.51 points in $\Robj$, so run-to-run variation includes platform effects. \Cref{tab:seeds} therefore reports the four-run envelope.

\subsection{Matched FNO: decision rule}\label{app:fno}
The comparison rule was fixed before any FNO training and is applied in both directions, per equal-seed pair. These are retrospective comparisons on cohort A. The rule has nine numerical checks and a visual veto:
\begin{itemize}
\item the winner improves mean $\Enear$ and $\Eobj$ by at least 5\% each;
\item it raises mean $\Robj$ by at least 0.02;
\item it regresses mean $\Evol$, $\Egrad$ and $\Epk$ by at most 2\% each;
\item it regresses the p90 of $\Enear$ and $\Eobj$ by at most 2\% each;
\item it wins jointly on $\Enear$ and $\Eobj$ in at least 9/12 geometries, with at most 0.01 recall loss.
\end{itemize}
Per seed, the reference-corrected surrogate lowers the FNO's $\Enear$ by 21.8--25.8\% and $\Eobj$ by 22.7--27.6\%, raises recall by 3.2--5.1 points, and wins jointly in 10--11 of 12 geometries. Inspection of 24 fixed-scale panels shows broader smooth residuals and stronger interface lobes for the FNO, and no artifact that reverses the numerical result. An earlier, unmatched direct FNO had $\Evol=13.5\%$ and $\Enear=17.0\%$ under the same composition; it used fewer training cases and updates.

\subsection{Origin of cohort B}
Cohort B was generated to test a later candidate: an equal-weight ensemble of a different surface-conditioned model, composed with the same reference. The reference-corrected predictor of this paper served as that test's sealed control. That candidate improved $\Eobj$ by 8.7\% but did not meet its own promotion rule, and it is not part of this paper. Here cohort B supplies prospective absolute scores of our predictor, a post hoc reconstructed reference contrast and regional and least-squares residual diagnostics, and retrospective reference-alone and rebased-surrogate baselines. The new geometries do not make these later-designed analyses prospective.

\end{document}